\documentclass{article}
\usepackage{graphicx}
\usepackage{dcolumn}
\usepackage{bm}
\usepackage{amsmath}
\usepackage{amssymb}
\usepackage[usenames]{color}
\usepackage[a4paper]{geometry}
\newcommand{\mbf}{\mathbf}

\begin{document}


\title{Micro-tearing modes in the Mega Ampere Spherical Tokamak}

\author{D. J. Applegate$^1$, C. M. Roach$^2$, J. W. Connor$^2$, S. C. Cowley$^3$, W. Dorland$^4$, R. J. Hastie$^2$, \\ and N. Joiner$^5$}

\maketitle

\noindent \small{$^1$ Department of Physics, University of York, UK.\\}
\small{$^2$ EURATOM/UKAEA  Fusion Association, Culham Science Centre, UK.\\}
\small{$^3$ Department of Physics and Astronomy, UCLA, USA.\\}
\small{$^4$ Department of Physics, University of Maryland, USA.\\}
\small{$^5$ Department of Physics and Engineering Physics, University of Saskatewan, Canada.}

\begin{abstract}
Recent gyrokinetic stability calculations have revealed that the spherical tokamak is susceptible to tearing parity instabilities with length scales of a few ion Larmor radii perpendicular to the magnetic field lines.  Here we investigate this `micro-tearing' mode in greater detail to uncover its key characteristics, and compare it with existing theoretical models of the phenomenon.  This has been accomplished using a full numerical solution of the linear gyrokinetic-Maxwell equations.  Importantly, the instability is found to be driven by the free energy in the electron temperature gradient as described in the literature.  However, our calculations suggest it is not substantially affected by either of the destabilising mechanisms proposed in previous theoretical models.   Instead the instability is destabilised by interactions with magnetic drifts, and the electrostatic potential.  Further calculations reveal that the mode is not significantly destabilised by the flux surface shaping or the large trapped particle fraction present in the spherical tokamak.  Its prevalence in spherical tokamak plasmas is primarily due to the higher value of plasma $\beta$, and the enhanced magnetic drifts  due to the smaller radius of curvature.
\end{abstract}
\normalsize

\section{Introduction}

The linear tearing instability, described by Furth, Killeen, and Rosenbluth \cite{killeen}, progresses by relaxing a sheared magnetic field into a lower energy magnetic island structure.  Importantly, the energy released from this process decreases as a function of mode number, which prevents the formation of shorter wavelength tearing instabilities.  However, later calculations by Hazeltine \emph{et al} \cite{haz}, demonstrated that the linear tearing instability could also be driven by an electron temperature gradient in slab geometry.  This new mechanism may enable the tearing instability to exist at higher mode numbers, giving rise to the term `micro-tearing mode.'  

For a given toroidal mode number $n$, the distance between adjacent rational surfaces is $\delta r=\frac{1}{nq'}$, where $q'$ is the radial derivative of the safety factor $q$.  Consequently, the chains of islands formed by micro-tearing instabilities are in close proximity, and may overlap producing a stochastic field line structure.  When this occurs, the electrons are no longer confined to equilibrium flux surfaces, and are able to move radially by moving along the perturbed field lines.  This process can produce significant electron heat transport, which could explain the high ratios of $T_i/T_e$ measured in a number of spherical tokamak (ST) plasmas  (including those from START and NSTX \cite{roachtransport,NSTXconf}).  

The drive mechanism for this micro-tearing instability can be explained in terms of the parallel thermal force \cite{has2}.  This force is due to the different frictional forces experienced by electrons travelling in opposite directions along a parallel temperature gradient, and has a magnitude given by,

\begin{equation}
F_{th} \sim -n_e \frac{\mbf{B}.\nabla T_e}{|\mbf B|}
\label{thermalforce}
\end{equation}

\noindent where $n_e$ and $T_e$ are the electron density and temperature respectively. The parallel current resulting from this force will produce a magnetic field perturbation $\mbf{\tilde B}$.  Instability then arises if this field perturbation aligns with an equilibrium electron temperature gradient $\nabla T_e$.   The later paper by Hassam \cite{has2}, uses a second order Chapman-Enskog expansion of the fluid equations to recover a similar solution.  This work notes that the time dependent thermal force is also important to the drive mechanism of the micro-tearing instability. (This time-dependent force is not present in first order Chapman-Enskog fluid theories, so the Braginskii equations \cite{brag} are unable to recover the instability).

Drake and Lee \cite{drake} expand on the work of Hazeltine \emph{et al} \cite{haz}, by carefully defining three collisional limits in which the slab/cylindrical micro-tearing instability can be analytically treated: the collisional regime, where $\bar \nu_{e} >> \omega$; the semi-collisional regime, defined by $\bar \nu_{e} > \omega$ and  $\bar \nu_{e} > k_\parallel^2 v_{th_e}^2/\omega $ (nb. perturbations take the form $\tilde \xi=\hat \xi e^{i(m\theta-n\phi-\omega t}$ such that $k_\parallel=(m-nq)/Rq$); and finally the collisionless regime where $\bar \nu_{e} << \omega$.  In these expressions, $v_{th_e}$ is the thermal velocity of electrons, $\omega$ is the mode frequency, $k_\parallel$ is the wavenumber of the mode parallel to the equilibrium field lines, and $\bar \nu_{e}= \frac{4\pi n_e e^4 \mathrm{ln}\lambda}{(2T_e)^{3/2}m_e^{1/2}}$ determines the collision rate experienced by electrons.  Hot tokamak plasmas typically have $\bar \nu_e \leq \omega^*_e$ (where $\omega^*_e = -k_yT_e/(eBn)dn/dr$ is the electron diamagnetic frequency), which best agrees with the collisionless limit of Drake and Lee.  Importantly the thermal force vanishes in this limit, meaning the collisionless \emph{micro-tearing} mode is stable in slab geometry \cite{drake,cowley} suggesting the micro-tearing mode may not be important in fusion plasmas such as ITER.

However, the paper by Catto and Rosenbluth \cite{catto} demonstrates another drive mechanism for the tearing instability in the ITER relevant banana regime  ($\bar \nu_e< \omega_{be}$ where $\omega_{be}$ is the electron bounce frequency).   In this limit electrons close to the trapped-passing boundary can easily scatter between the trapped and passing populations.  This increases the effective collision rate in a layer around the trapped-passing boundary and allows a destabilising current to flow.  Catto and Rosenbluth \cite{catto} have shown this can destabilise the tearing mode in the regime $\bar \nu_{e}/\epsilon<\omega$.  In similar spirit,  Connor \emph{et al} \cite{connor} show the trapped particle driven mode is stable when $\bar \nu_{e}/\epsilon>\omega>\bar \nu_{e}$. 
 
Importantly, gyrokinetic simulations have identified the micro-tearing mode in a number of ST H-mode plasmas \cite{roach,colineps,redi,howard}.  Here we build on these previous calculations with a closer investigation of the instability.  In particular, the key characteristics of the instability are uncovered, and compared with existing theoretical models of the phenomenon.  Section \ref{model} of this paper describes the equilibrium model we have considered, and also the system of equations employed in the study of the micro-tearing mode.  In Section \ref{results} the key features of the instability are discussed, including the sensitivity to temperature/density gradients, beta, collisionality, geometry, trapped particles, and also the importance of the electrostatic potential and the ion response.   These results confirm the micro-tearing modes found in MAST are not well described by the existing literature, and reveal the processes that are crucial for instability.  Our conclusions are given in Section \ref{conclusion}.

\section{Plasma Model} \label{model}

\begin{figure}[tbh]
\centering
\includegraphics[angle=270, width=6cm, totalheight=4.7cm,trim=0 0 -20 0,clip]{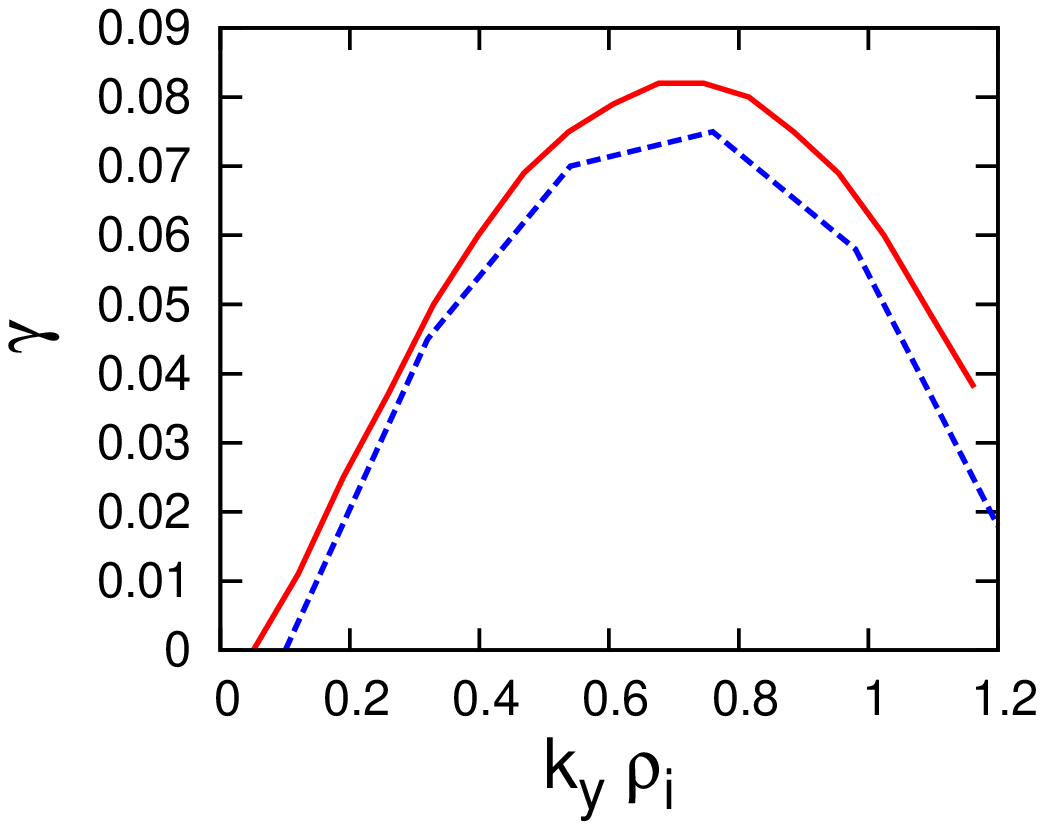}
\includegraphics[angle=270, width=6cm, totalheight=4.7cm,trim=0 0 -20 0,clip]{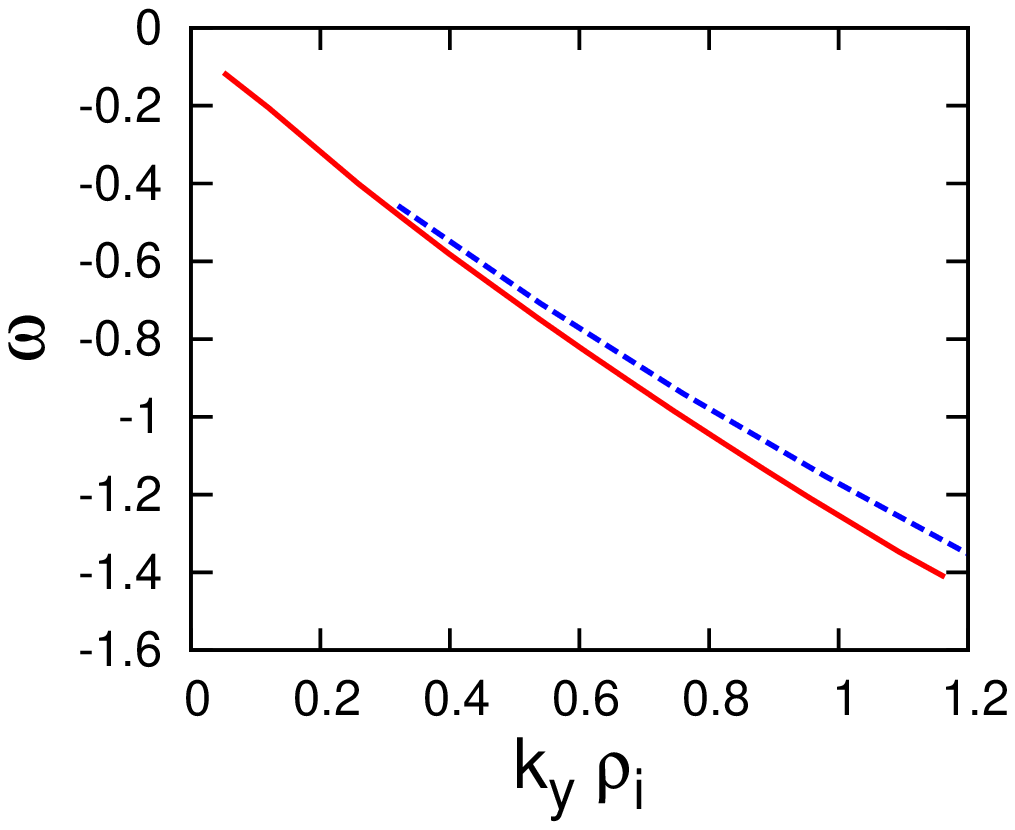}		
\caption{Growth rate ($\gamma$) and real frequency ($\omega$) of micro-tearing modes on the $\psi_n=0.4$ surface (both are normalised to $v_{th_i}/a$).  These are plotted as functions of the normalised wave number $k_y\rho_i$ (see Appendix A for the definition of the $x$ and $y$ coordinates used in GS2).  The solid line represents the two species model, while the dashed line represents the five species model.}
\label{fig:species_growths}
\end{figure}

The plasma considered is MAST discharge \#6252, which is a well diagnosed Elmy H-mode plasma with neutral beam heating.   An equilibrium reconstruction for this plasma has been calculated during a steady level of the plasma current, and just before the neutral beams were switched off. (The key plasma parameters are listed in Table \ref{tab:6252_global}, and a detailed discussion of the equilibrium calculation has been given in a previous publication \cite{roach}).  In particular we study the $\psi_n=0.4$ poloidal flux surface of this equilibrium ($\psi_n= \frac{\psi-\psi_{min}}{\psi_{max}-\psi_{min}}$ where $\psi$ is the poloidal flux), since this was found to be highly susceptible to the micro-tearing mode. (The important plasma parameters for this surface are given in Table \ref{tab:species}).    A significant difference between the plasma model employed here and that used in previous work \cite{roach}, is the reduction in the number of kinetically modelled species from five to just two (main ions and electrons).  This has significantly improved the tractability of our calculations without significantly affecting the results (see Figure \ref{fig:species_growths}). 

\begin{table}[htb]
\begin{center}
\begin{tabular}{|c|c|c|c|c|c|c|} \hline
$B_0$ (T) & $I_p$ (MA) & $R$ (m) & $a$ (m) & $R_{\rm mag}$ (m) & 
 $P_{\rm beam}$ (MW) & $Z_{\rm eff}$  \\
0.458 & 0.738 & 0.816 & 0.558 & 0.901 &  1.67 & 1.5  \\
\hline
\end{tabular}
\caption{Global parameters for the MAST equilibrium at $t=0.265s$ in MAST discharge \#6252.  Here $B_0$ is the toroidal magnetic field on the magnetic axis, $I_p$ is the plasma current, $R$ and $a$ are respectively the major and minor radius of the last closed flux surface, $R_{mag}$ is the major radius of the magnetic axis. $P_{beam}$ is the applied neutral beam heating, and finally  $Z_{eff}=\sum_{s} Q_{s}^2n_s/n_e$ (where $s$ labels the ion species and $Q_s$ is the electric charge of the ion species.).}
\label{tab:6252_global}  
\end{center}
\end{table}

\begin{table}[htb]
\centering
\begin{tabular}{|c|c||c|c|}
\hline
$T_e$ 			& 592 $eV$   												& $\nu_e$ 	&   0.429   \\
$T_i$ 			& 611 $eV$ 												  & $\nu_i$	 	&   0.0125  \\
$a/L_{T_e}$ & 2.04   													& $\beta$ 	&   0.1  \\
$a/L_{T_i}$ & 2.04   													& $q$  			&		1.346  \\
$n$ 				& 4.36 $\times 10^{19} m^{-3}$  	& $\hat s$	&		0.286   \\
$a/L_n$   	& -0.177    												& $r$  			&		0.311m   \\
$P$  				& 838 Pa  													& $\rho_i$	&		7.64 mm \\
$a/L_p$  		& 1.87   													&   				&           \\
\hline
\end{tabular}
\caption{Important parameters for the poloidal flux surface $\psi_n=0.4$.  $T_e$ and $T_i$ are the electron and ion temperatures, $P$ is the plasma pressure, while $n$ gives the number density for both ions and electrons. The normalised gradient of a quantity $x$ is defined as $a/L_x=\frac{-1}{x}\frac{d x}{d\psi_n}$, while $\hat s=\frac{\psi_n}{q} \frac{d q}{d\psi_n}$.  The parameters $\bar \nu_i=\bar \nu_e Q_i^2 Z_{eff} \left(\frac{m_e}{m_i}\right)^{1/2} \left(\frac{T_e}{T_i}\right)^{3/2}$, and $\bar \nu_e=\frac{4\pi n_e e^4 \mathrm{ln}\lambda}{(2T_e)^{3/2}m_e^{1/2}}$, determine the collision rates of ions and electrons (respectively) and are both normalised to $v_{th_i}/a$.  The quantity $r$ is the half diameter of the flux surface across its midplane, while $\rho_i$ is the Larmor radius of thermal ions.} 
\label{tab:species}
\end{table}

The linear gyrokinetic-Maxwell equations \cite{gyrokinetics,catto1} have been employed to study the micro-tearing instability present in this equilibrium. These equations provide a kinetic description for the plasma in the limit,

\begin{eqnarray}
\frac{\omega}{\Omega} \approx \frac{k_{\parallel}}{k_{\perp}} \approx \frac{\rho}{L} \approx \delta \ll 1 &&  k_\perp \rho \sim  1
\end{eqnarray}

\noindent where $\Omega$ is the gyro-frequency, $\omega$ is the oscillation frequency of the field perturbations, $k_\perp$ and $k_\parallel$ are the perpendicular and parallel wavenumbers, and $L$ is the scale length of variations in the equilibrium distribution function.   The linear gyrokinetic equation for a Fourier mode $\mbf{k_\perp}$ is given by the expression below.

\begin{eqnarray}
\frac{\partial \hat g_{s,\mbf{k_\perp}}}{\partial t}   + \mbf {v_\parallel}.\frac{\partial \hat g_{s,\mbf{k_\perp}}}{\partial \mbf R} + \frac{1}{\Omega}\mathbf{\hat z}\times (\mu \frac{\partial B}{\partial \mbf r} + v_\parallel^2 \mathbf{\hat z}.\frac{\partial B}{ \partial \mbf r}).\frac{\partial \hat g_{s,\mbf{k_\perp}}}{\partial \mathbf{R}}  -
\left(\frac {Q_s F_{0,s}}{T_s} \frac{\partial}{\partial t} + i\frac{\partial F_{0,s}}{\partial \mbf R} \times \frac{\mathbf{\hat z}}{B}. \mathbf {k_\perp} \right) \zeta_{\mbf{k_\perp}} = \mathcal{C} [\hat g_{s,\mbf{k_\perp}}]
\label{nonlineareik}
\end{eqnarray}

\noindent where $\mathbf{R}=\mathbf{r}-\frac{\mathbf{\hat{z}\times v}}{\Omega_s}$ is the guiding centre coordinate;  the nonadiabatic distribution function is given by $g_s=\tilde f_s + \frac{Q_s \phi F_{0,s}}{T_s}$ (where $Q_s$ is the particle charge, $\tilde \phi$ is the electrostatic potential,  $\tilde f_s$ is the perturbed distribution function, and $F_{0,s}$ is the Maxwell-Boltzmann distribution function); finally  $\zeta_{\mbf{k_\perp}}=$  $\left[ \hat \phi_{\mbf{k_\perp}} J_0(z)\right.$ $- v_\parallel \hat A_{\parallel,\mbf{k_\perp}} J_0(z)+$ $\frac {v_\perp}{\left|\mbf{k_\perp}\right|}$ $\left. \hat  B_{\parallel,\mbf{k_\perp}} J_1(z) \right]$ where $z=k_\perp v_\perp/\Omega$.   Particle collisions are modelled using a Lorentz collision operator,

\begin{eqnarray}
\mathcal{C}[\hat g_{s,k_\perp}] = \frac{\nu_s(v)}{2}\left[\left.\frac{\partial}{\partial \xi}\right|_{\mathbf{R}}(1-\xi^2)\left.\frac{\partial g_{s,k_\perp}}{\partial \xi}\right|_{\mathbf{R}}\right. 
\left.-\frac{v^2(1+\xi^2)}{2\Omega_s^2}k_\perp^2 g_{s,k_\perp} \right]  \label{eqn:alexcol}
\end{eqnarray}

\noindent where $\nu$ is the particle collision rate, $\xi=v_\parallel/v$ is the pitch angle variable, and $|_{\mbf R}$ denotes a partial derivative at constant $\mbf R$.  This operator calculates the pitch angle scattering of particles due to collisions, which enables an accurate model of trapped particle interactions. It also conserves energy and particle numbers.  However it does not model the energy scattering involved in collisions, and does not include momentum conservation.  The diffusive term proportional to $k_\perp^2$ is found to have little influence on the linear micro-tearing mode \cite{colineps}.  However, preliminary work suggests it may have a more important effect in nonlinear simulations of the instability \cite{applegate}.  Both ion-electron collisions and electron-electron collisions are included in the electron collision rate $\nu_e(v)$, while only ion-ion collisions are included in the ion collision rate $\nu_i(v)$. However, it is found that electron-ion collisions are by far the most important of these collisional processes.

The gyrokinetic-Maxwell equations have been solved numerically using the initial value code GS2 \cite{kotgs2}.  This code assumes the equilibrium temperature and density scale lengths are constant across the radial simulation domain, and since the radial simulation domain is generally quite small for the micro-tearing instability in MAST, this approximation is reasonable (see reference \cite{roach}). Two further assumptions have also been made to improve the numerical tractability of these calculations .  First the parallel magnetic perturbations ($\mbf{\tilde B}_\parallel=\nabla \times \mbf{A_\perp}$) have been neglected, which, despite the high plasma $\beta$, have little effect on the micro-tearing mode (see Section \ref{betaeffect}).  Second, the plasma has been approximated as quasi-neutral. (Results obtained using the quasi-neutrality condition were found to be in good agreement with those obtained from Poisson's equation.)

\section{Results} \label{results}

\begin{figure*}[htb]
\centering
\includegraphics[angle=270, width=5.8cm, totalheight=5.15cm,trim=0 20 0 20,clip]{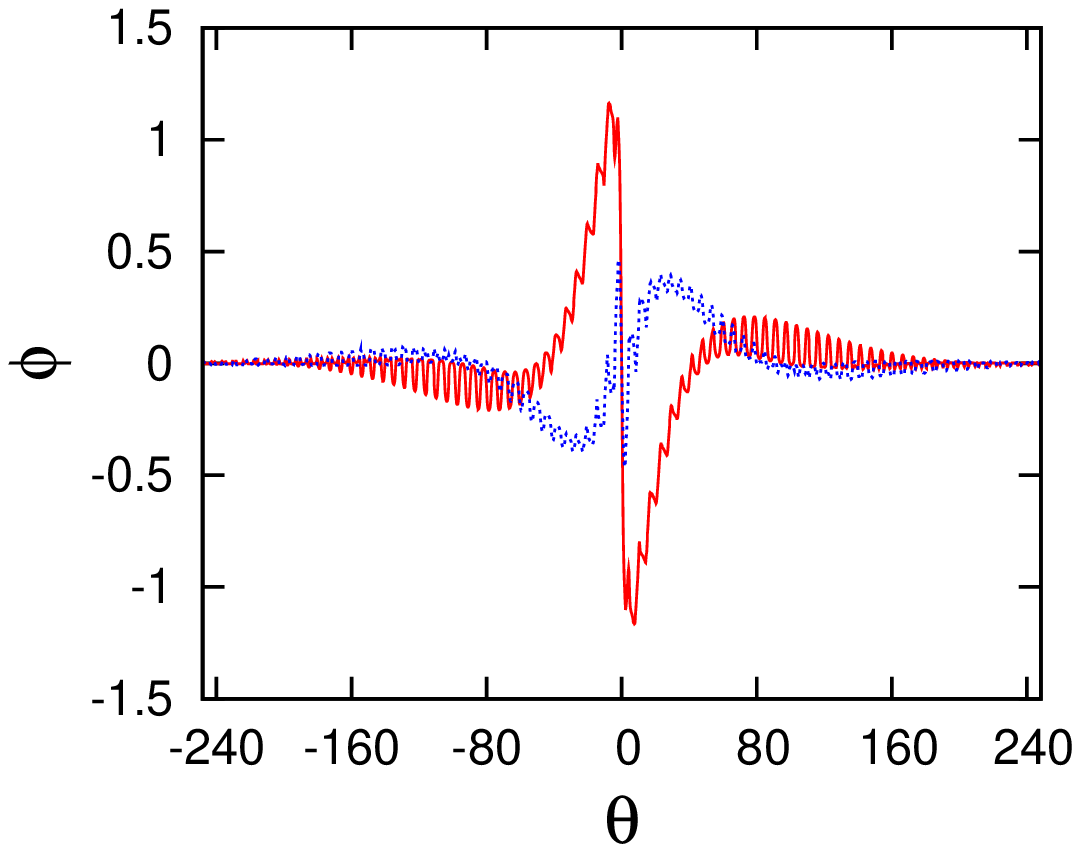}
\includegraphics[angle=270, width=5.8cm, totalheight=5.15cm,trim=0 20 0 20,clip]{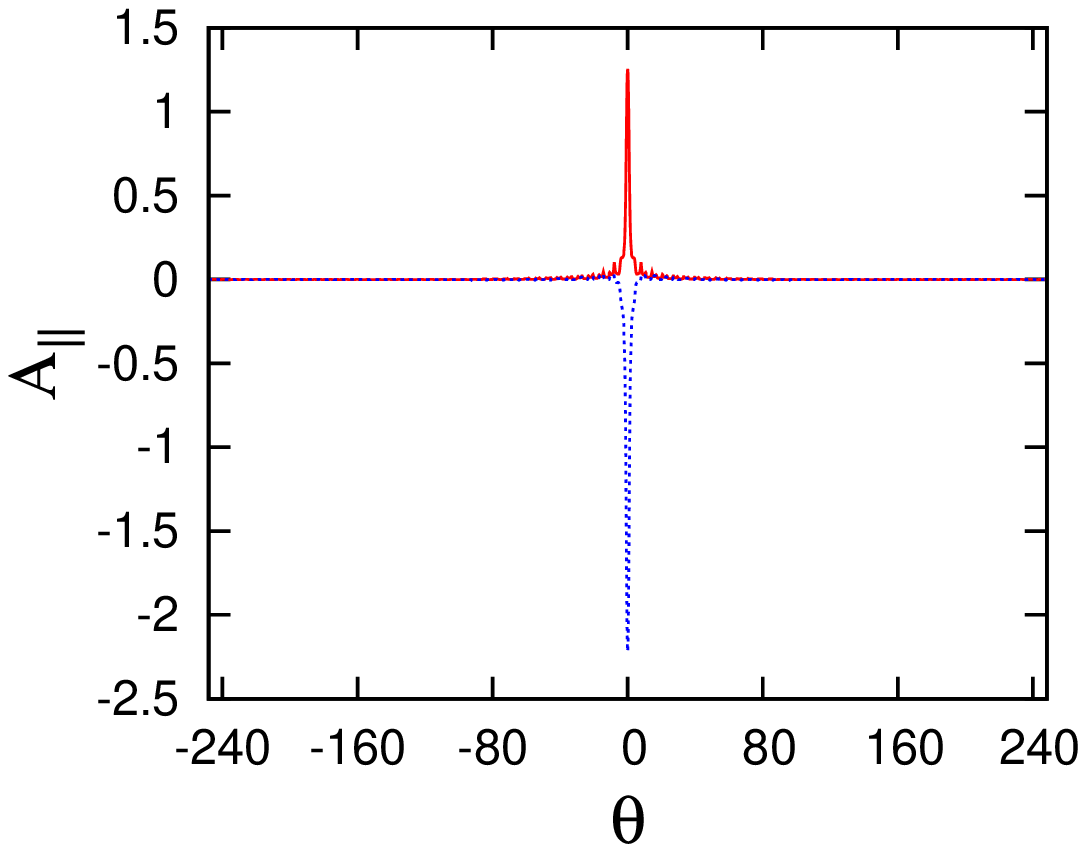}		
\includegraphics[angle=270, width=5.8cm, totalheight=5.15cm,trim=0 20 0 20,clip]{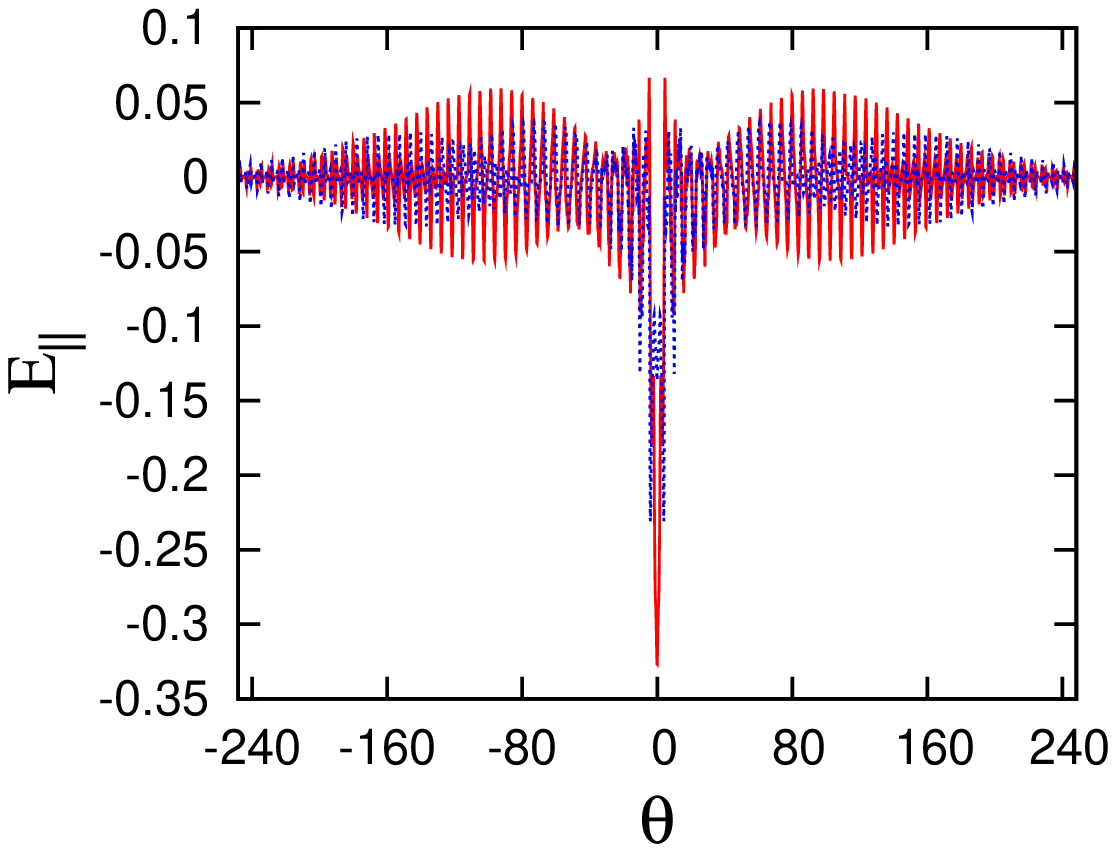}
\caption{Eigenfunctions for the electrostatic potential $\phi$, the parallel vector potential $A_\parallel$, and the parallel electric field $E_\parallel$ plotted as functions of the ballooning angle $\theta$.  The real part of each eigenfunction is represented by a solid line, whilst the imaginary parts are represented by dotted lines.}
\label{fig:eigenfunctions}
\end{figure*}

Figure \ref{fig:eigenfunctions} is a plot of the parallel electric field $\tilde E_\parallel$, the electrostatic potential $\tilde \phi$, and the parallel vector potential $\tilde A_\parallel$ resulting from the micro-tearing instability at $k_y\rho_i=0.5$.  Each quantity is plotted as a  function of the ballooning angle $\theta$ (which labels the distance along a magnetic field line using the poloidal angle).  Importantly, the electrostatic potential and parallel electric field are highly extended along the field lines due to the rapid parallel motion of electrons.  The parallel vector potential is confined to a much smaller range in $\theta$, which, in real space, produces a radially extended function.  This corresponds to the ``constant $\psi$" approximation of linear tearing theory \cite{killeen}. 

The tearing nature of this instability becomes apparent when Poincar\'e sections are plotted of the perturbed magnetic field.  In GS2 this is achieved using a flux tube simulation domain (see reference \cite{beercoord} for details), which is a long thin tube aligned parallel to the magnetic field lines.  The parallel length extends from $\theta=[-\pi, \pi]$ (as opposed to the arbitrarily large domain of the ballooning representation), while the perpendicular dimensions are a few Larmor radii.  Parallel vector potential data from GS2 is then used to determine the path of selected magnetic field lines through the flux tube.  Once these paths have been established, a perpendicular slice is taken through the tube at constant poloidal angle $\theta$, and the points where the field lines intersect with this slice are plotted. 

\begin{figure} [tbh]
\centering
\includegraphics[angle=90,width=4.5cm, totalheight=7.0cm,trim=50  50  10 15,   clip]{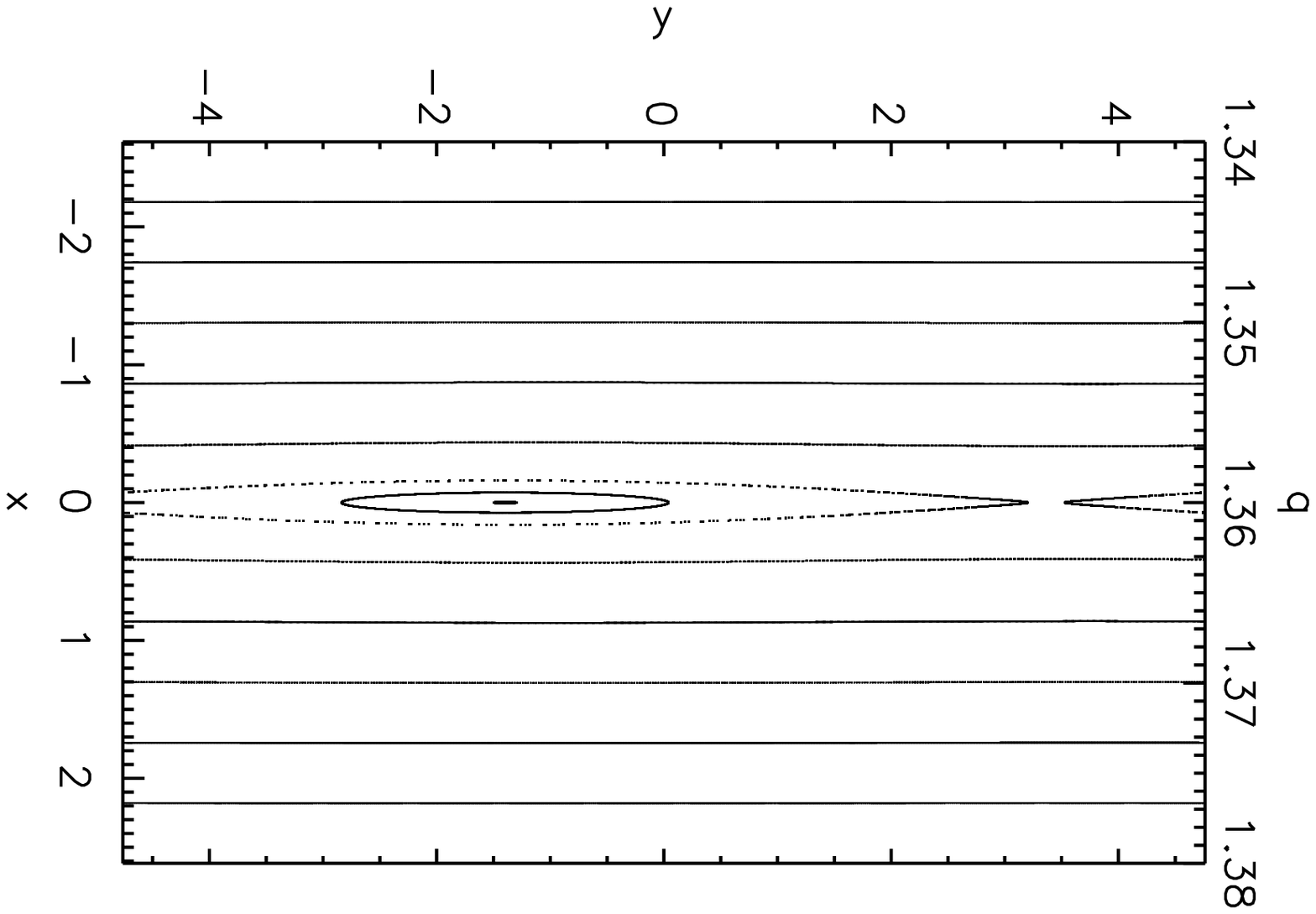} 
\includegraphics[angle=90,width=4.5cm, totalheight=7.0cm,trim=50  50  10 15,   clip]{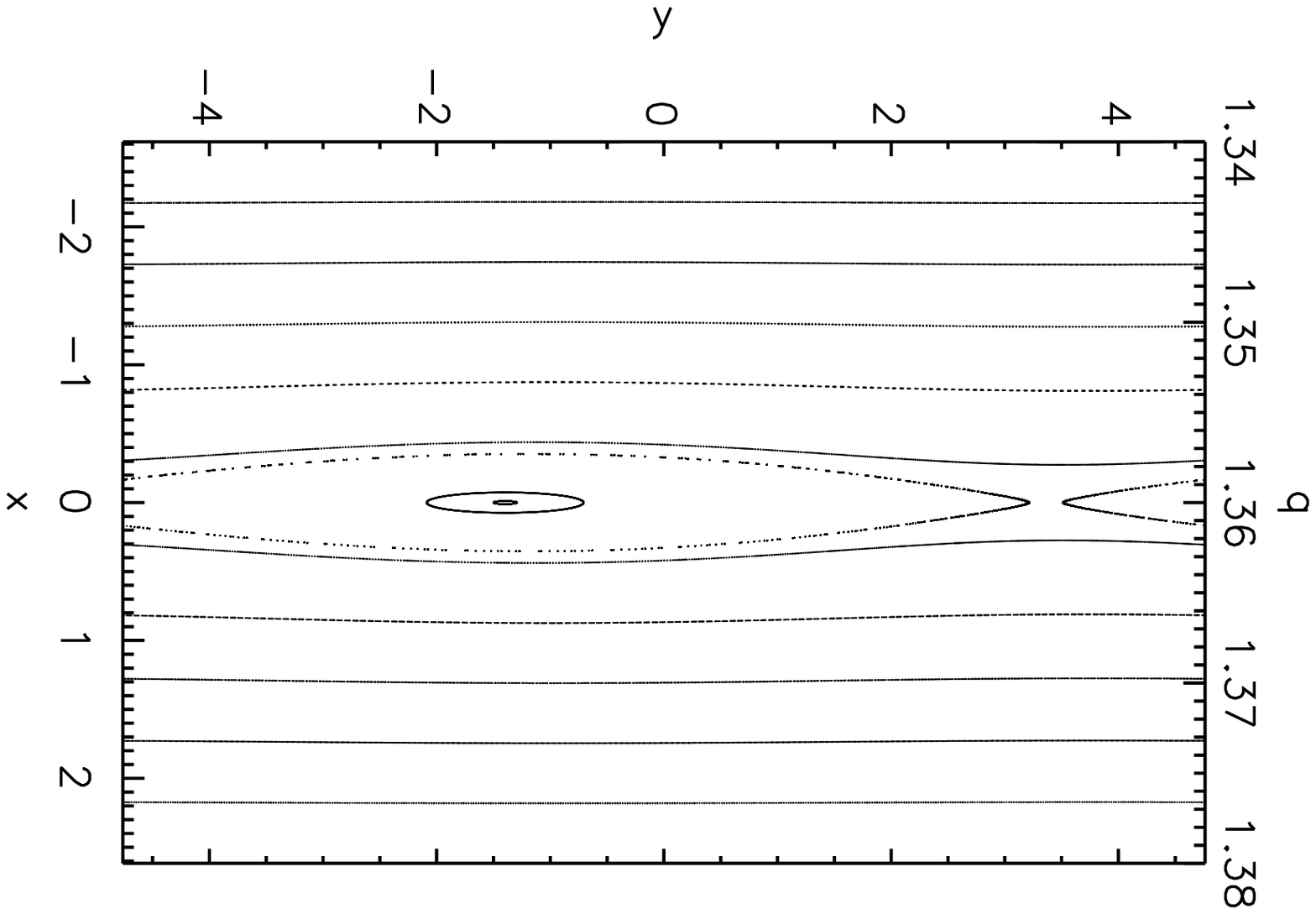}
\includegraphics[angle=90,width=4.5cm, totalheight=7.0cm,trim=50  50  10 15,   clip]{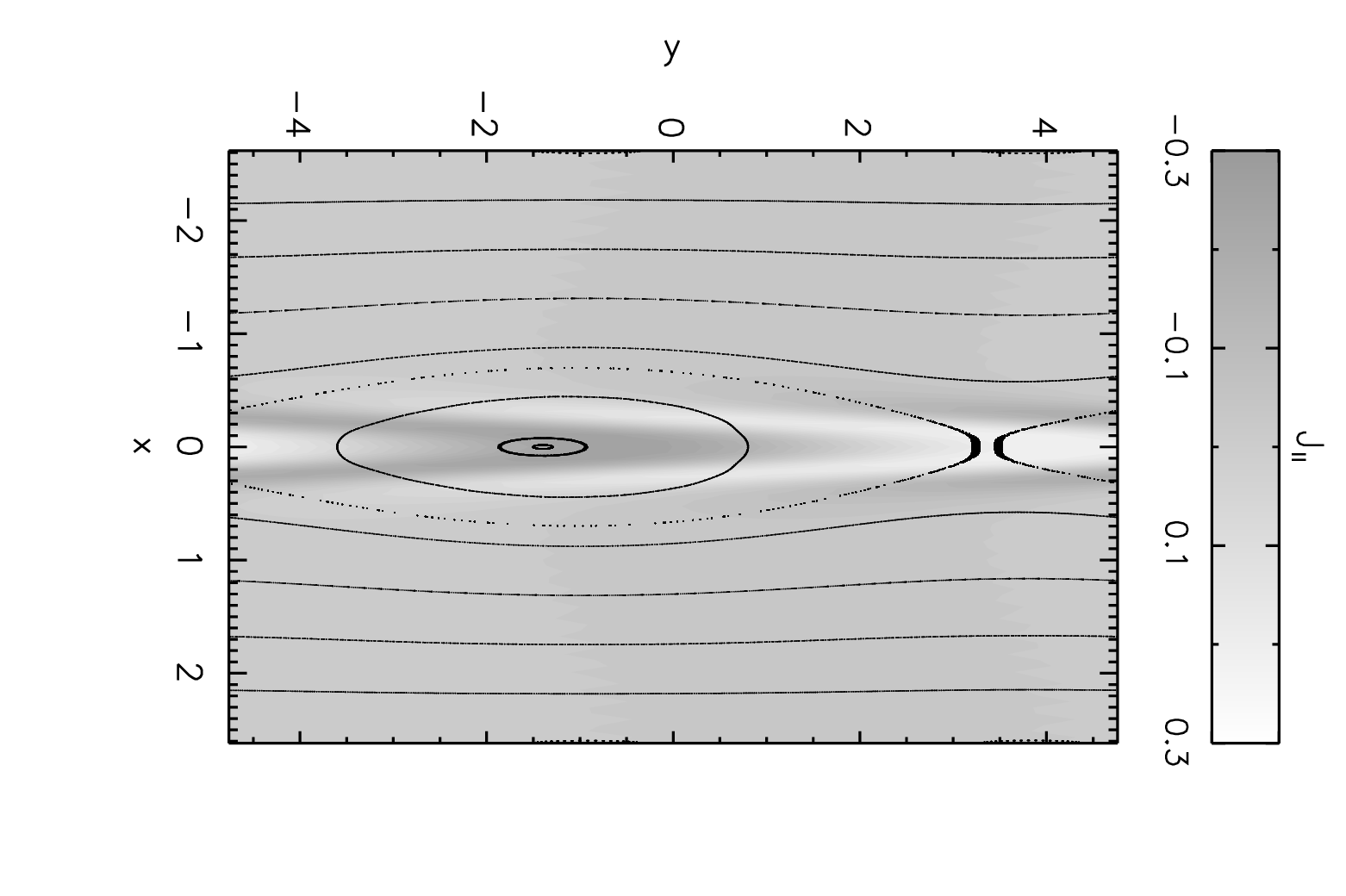}
\caption{Poincar\'e plots for the micro-tearing instability with $k_y\rho_{i}$=0.66 at $\theta=0$, where the $x$ and $y$ axis are given in units of $\rho_i$.  The amplitude of the perturbation increases from left to right, so that the middle plot has a magnitude five times greater than the left hand plot, and four times smaller than the right hand plot. The right hand plot also includes a contour diagram of the parallel current $\tilde J_\parallel$.}
\label{fig:poincare}
\end{figure}

The Poincar\'e plots in Figure \ref{fig:poincare} were produced using $\tilde A_\parallel$ data for a single micro-tearing mode with wavenumber $k_y\rho_{i}=0.66$.  Each plot is given on an $(x,y)$ grid, where $x$ and $y$ are the perpendicular coordinates in the flux tube, and are both normalised to the ion Larmor radius $\rho_i$. (The definitions of $x$ and $y$ are given in Appendix A).   A second horizontal axis labels the equivalent $q$ values for the $x$ axis.  Importantly, the field lines are tracked from the same starting points in each plot, while the amplitude of the magnetic field perturbation (due to the instability) increases from left to right. A magnetic island structure appears at the rational surface $q = \frac{34}{25}$ of each plot which increases in size and complexity as the amplitude of the perturbation increases.  The current plot in Figure \ref{fig:poincare} reveals the `O' point of the island is at the current minimum, whilst the `X' point of the separatrix is at the current maximum.  This current perturbation occurs in  a narrow region around the resonant surface, and corresponds to the current layer of linear tearing mode theory \cite{killeen}. 

\subsection{Drift-wave Characteristics of the Micro-tearing Mode} \label{drift}

\begin{figure} [htb]
\centering
\includegraphics[angle=270, width=6cm, totalheight=4.7cm,trim=0 0 -20 0,clip]{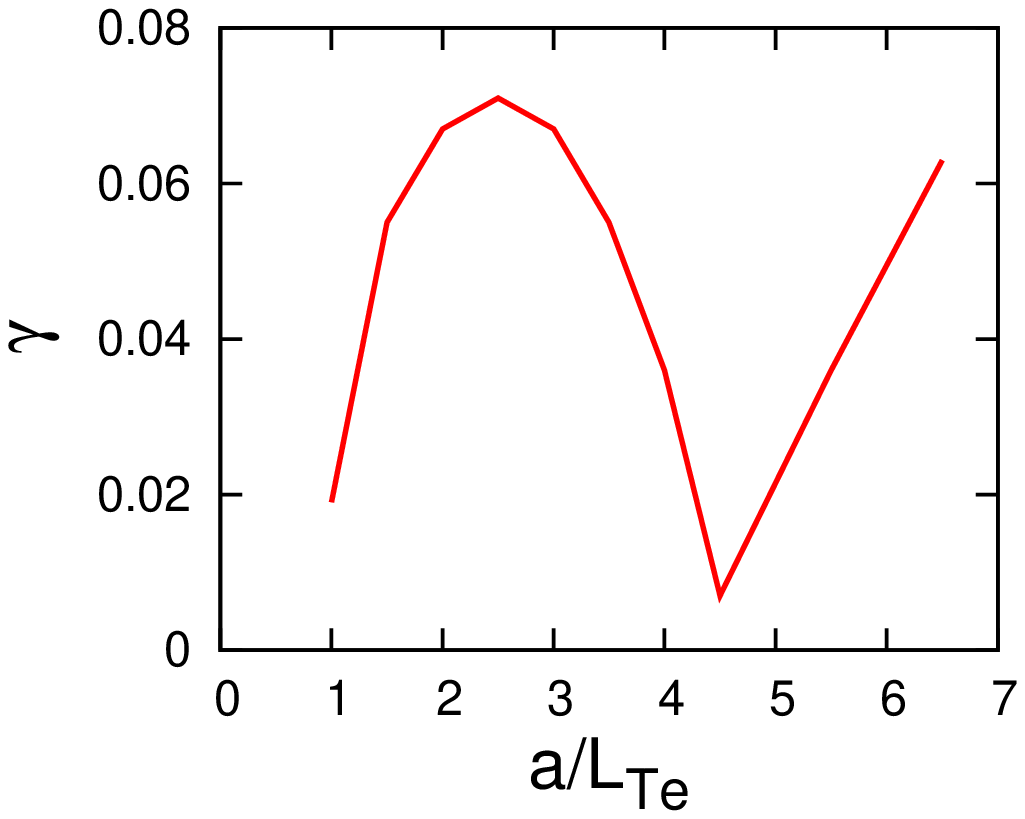}
\includegraphics[angle=270, width=6cm, totalheight=4.7cm,trim=0 0 -20 0,clip]{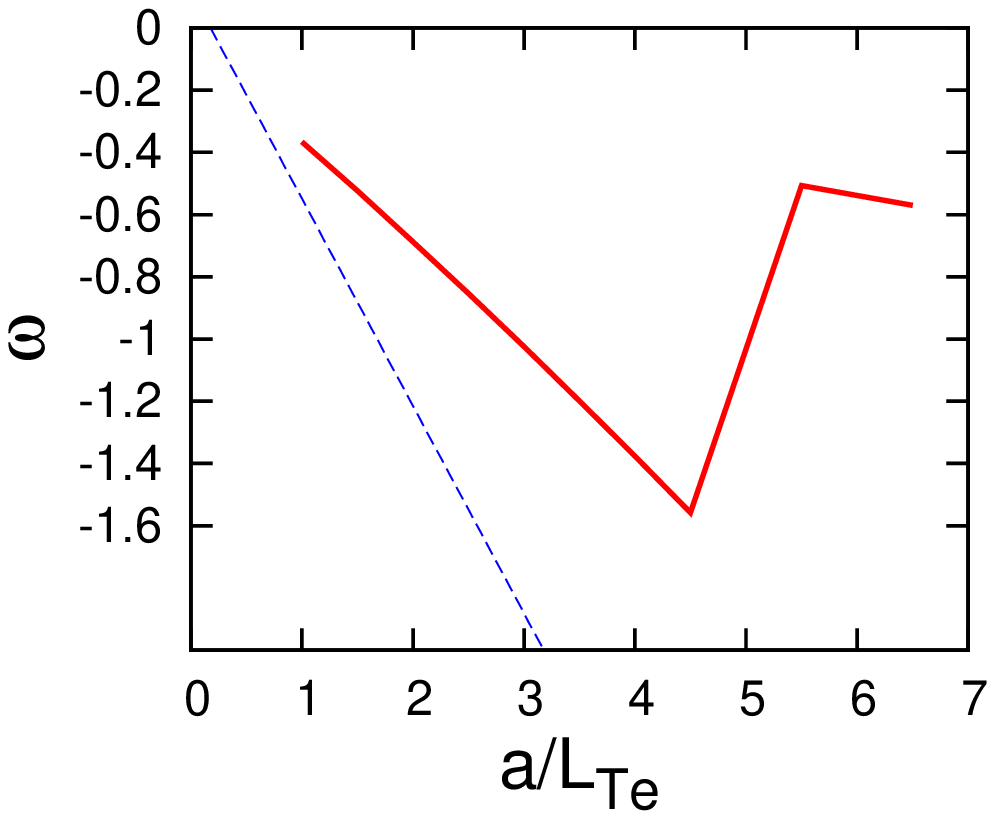}		
\caption{Growth rates and real frequency of micro-tearing modes as a function of the inverse electron temperature scale length $a/L_{T_e}$. For comparison, the dashed line in the real frequency plot represents the analytic prediction $\omega\sim \omega_e^*[1+\eta_e/2]$.  Note that the inverse electron density scale length was kept constant, so $\eta_e=\frac{L_n}{L_T}$ varies proportionally with $a/L_{T_e}$.}
\label{fig:density_temperature}
\end{figure}

The thermal force and trapped particle drive mechanisms for the micro-tearing instability both require a finite electron temperature gradient from which to draw free energy.  The left hand side of Figure \ref{fig:density_temperature} is a plot of growth rate as a function of $a/L_{Te}$ for the MAST micro-tearing instability with $k_y\rho_i = 0.5$. (Note, the ion temperature gradient was simultaneously altered to maintain a constant pressure gradient.)  The instability is clearly dependent on the presence of a finite electron temperature gradient, as predicted by Hazeltine \emph{et al} \cite{haz}.  Further calculations have shown the growth rate also varies with the electron density gradient, although a finite density gradient is not vital to the instability.

Previous theories of the micro-tearing mode often predict a linear relation between the equilibrium electron temperature and density gradients and the real frequency of the mode.  For example the paper by Catto and Rosenbluth \cite{catto} predicts $\omega\sim \omega_e^*[1+\eta_e/2]$ (where $\eta_e=\frac{L_{n_e}}{L_{T_e}}$), while numerical calculations by Gladd \emph{et al} \cite{gladd} produce real frequencies which roughly fit the expression, $\omega \sim \omega^*_e[1+\left[1+0.6\,\mathrm{ln}\left(\frac{\bar \nu_{e}}{\omega^*_e}\right)\right]\eta_e]$.  Both of these expressions have a qualitative agreement with calculations for the MAST instability (see Figure \ref{fig:density_temperature} for the dependence of the real frequency on the electron temperature gradient).   However, the frequency of the MAST instability also depends on the collision rate, which is closer to the predictions of Gladd \emph{et al}.  

Curiously there is a minimum in the growth rate of Figure \ref{fig:density_temperature} at $a/L_{Te} \approx 4.5$, which coincides with a discontinuity in the frequency plot.  This signifies a mode switching between two different micro-tearing instabilities, one which is unstable for $0<a/L_{Te}<4.5$ (which we will call mode \textbf{A}), and another which is unstable for $a/L_{Te}>4.5$ (which we will call mode \textbf{B}).  It is not obvious why mode \textbf{A} should be stabilised above $a/L_{Te} \approx 4.5$ since the temperature gradient provides the free energy for the instability. A plausible explanation could be the dependence of the real frequency on the electron temperature gradient.  If a resonance is important (for example with the magnetic drift frequency), then increasing $|\omega|$ may disturb this resonance, and cause mode \textbf{A} to be damped.   Interestingly, $|\omega|$ sharply decreases across $a/L_{Te}\sim4.5$.  This means mode \textbf{B} has a real frequency similar to the real frequency of mode \textbf{A} at its most unstable, and supports the idea that a lower $|\omega|$ produces a destabilising resonance in the plasma.

\subsection{The Role of Collisions}

It was discussed earlier that Drake and Lee \cite{drake} have defined three specific categories of tearing mode depending on the level of collisionality.  However, it is not obvious which of these three categories the MAST instability belongs to. Consider Figure \ref{fig:collisions4}, which shows the growth rate of the MAST micro-tearing instability as a function of collisionality. The instability is not collisionless, since the growth rate is clearly dependent on $\bar \nu_{e}$, and instability requires finite $\bar \nu_e$ (Figure \ref{fig:collisions4}).  It is also not collisional or semi-collisional, since the collision rate is too low:  $0.05 \lesssim \bar \nu_{e}/\omega \lesssim 2$.  In fact the instability seems to lie somewhere between the collisionless and semi-collisional regimes.  Importantly, the MAST instability is unstable for  $\bar \nu_{e}\sim \omega$.  This regime has been addressed by Gladd \emph{et al} \cite{gladd} and D'Ippolitto \emph{et al} \cite{dip} using a modified Krook collision operator \cite{will,krook}.  Conversely, at lower collisionalities, the MAST instability is closer to the limit treated in the trapped particle calculations of Catto and Rosenbluth \cite{catto}, where $\omega>\bar \nu_{e}/\epsilon$.  

\subsubsection{The Energy Dependence of the Collision Operator}

Importantly, the thermal force drive mechanism of Hazeltine \emph{et al} \cite{haz}, and the trapped particle drive mechanism of Catto and Rosenbluth \cite{catto}, both require an energy dependent collision operator.   Consequently we have tested whether the MAST instability is sensitive to the energy dependence of collisions.  This test involved replacing the velocity/energy dependent collision rate of the electrons with the collision rate of electrons moving at their thermal velocity,

\begin{equation}
	\nu_e(v)\rightarrow \nu_e(v_{th_e})  \label{eqn:energyind}
\end{equation}

\begin{figure} [tbh]
\centering
\includegraphics[angle=270, width=6cm, totalheight=4.7cm,trim=0 0 -20 0,clip]{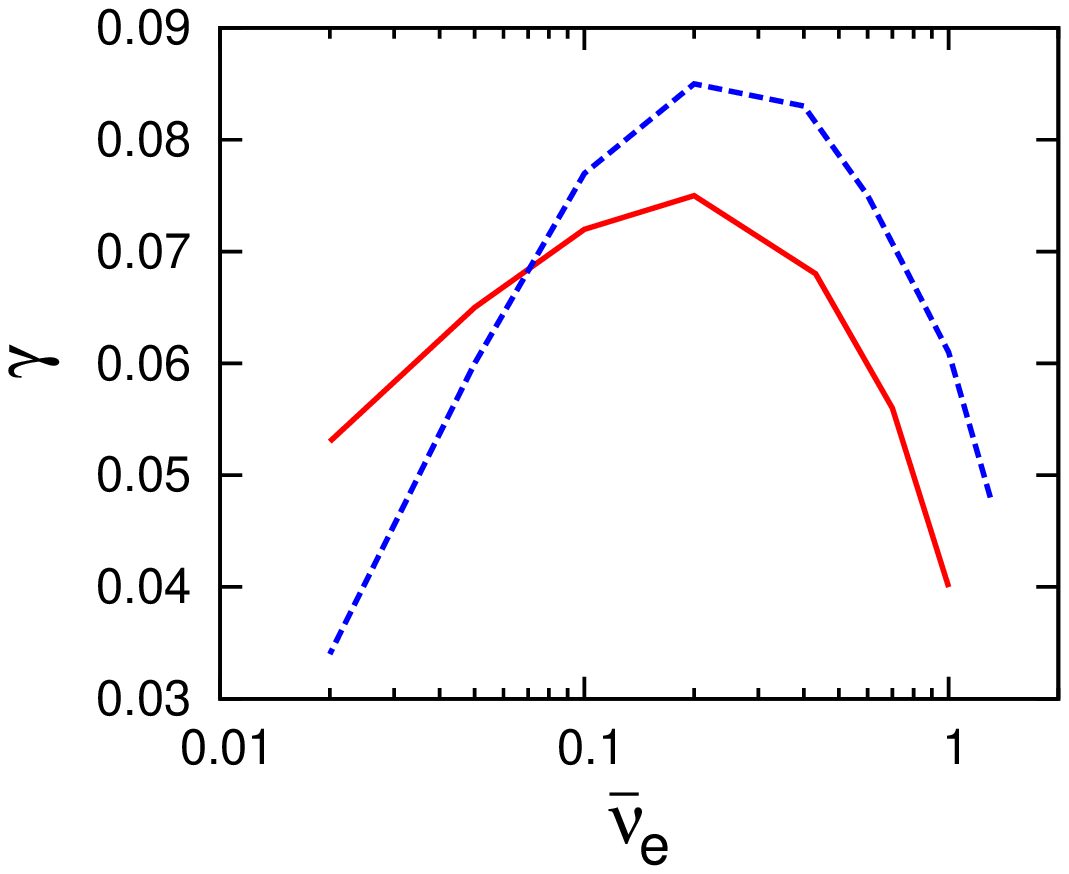}
\includegraphics[angle=270, width=6cm, totalheight=4.7cm,trim=0 0 -20 0,clip]{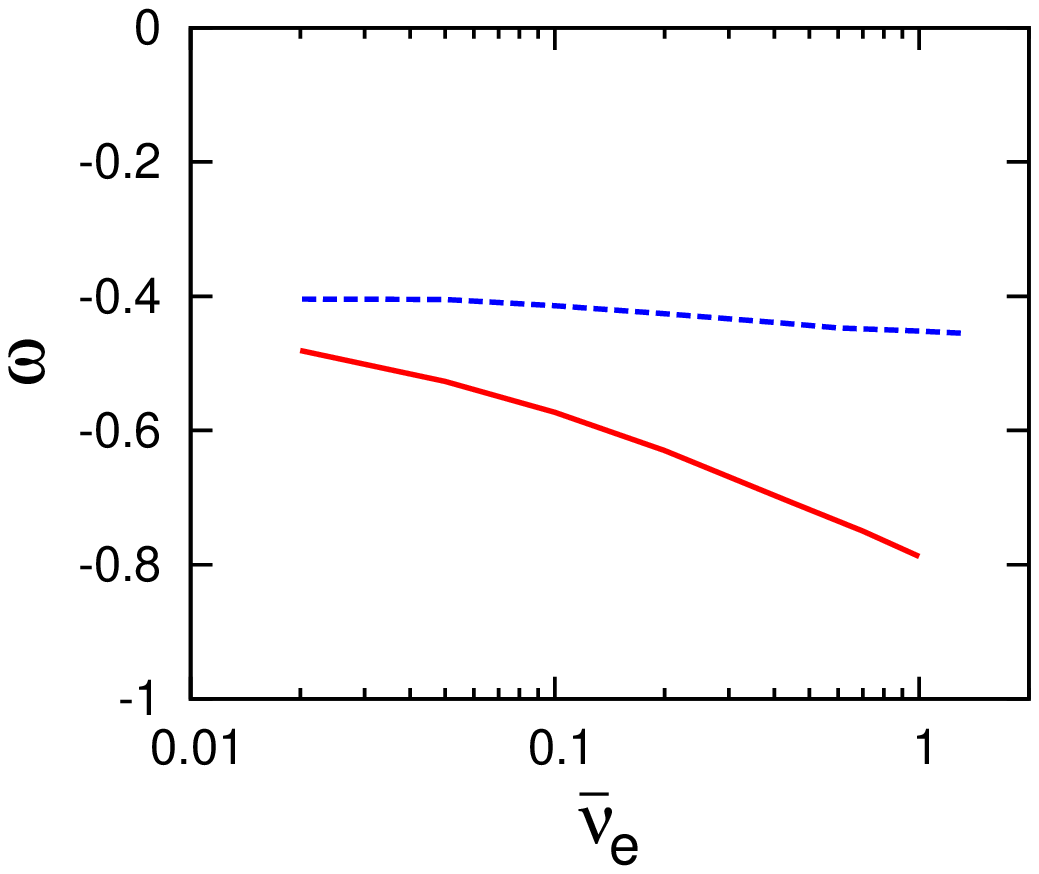}		
\caption{Plots of growth rate and real frequency for micro-tearing instabilities as a function of electron collisionality $\bar \nu_{e}$, for two different collision operators.  The solid lines represent the energy dependent collision operator, while the dashed lines represent the energy independent collision operator from equation (\ref{eqn:energyind}).}
\label{fig:collisions4}
\end{figure}

Figure \ref{fig:collisions4} is a plot of growth rate and real frequency versus the electron collisionality $\bar \nu_{e}$, using both the energy dependent collision operator and the energy independent collision operator.  It is clear the instability is not strongly affected by the change in collision operator, and thus cannot be strongly dependent on either of the drive mechanisms mentioned in the literature.

However, in the absence of magnetic drifts (see Section \ref{drifts}) energy dependent collisions become more important to the MAST instability. This makes contact with the numerical calculations of Gladd \emph{et al} \cite{gladd} and D'Ippolitto \emph{et al} \cite{dip}, who find a micro-tearing instability in slab geometry, which is reliant on energy dependent collisions, and occurs in the regime $\bar \nu_e \sim \omega$.  

\subsection{The Role of the Ion Response}

\begin{figure} [tbh]
\centering
\includegraphics[angle=270, width=6cm, totalheight=4.7cm,trim=0 0 -20 0,clip]{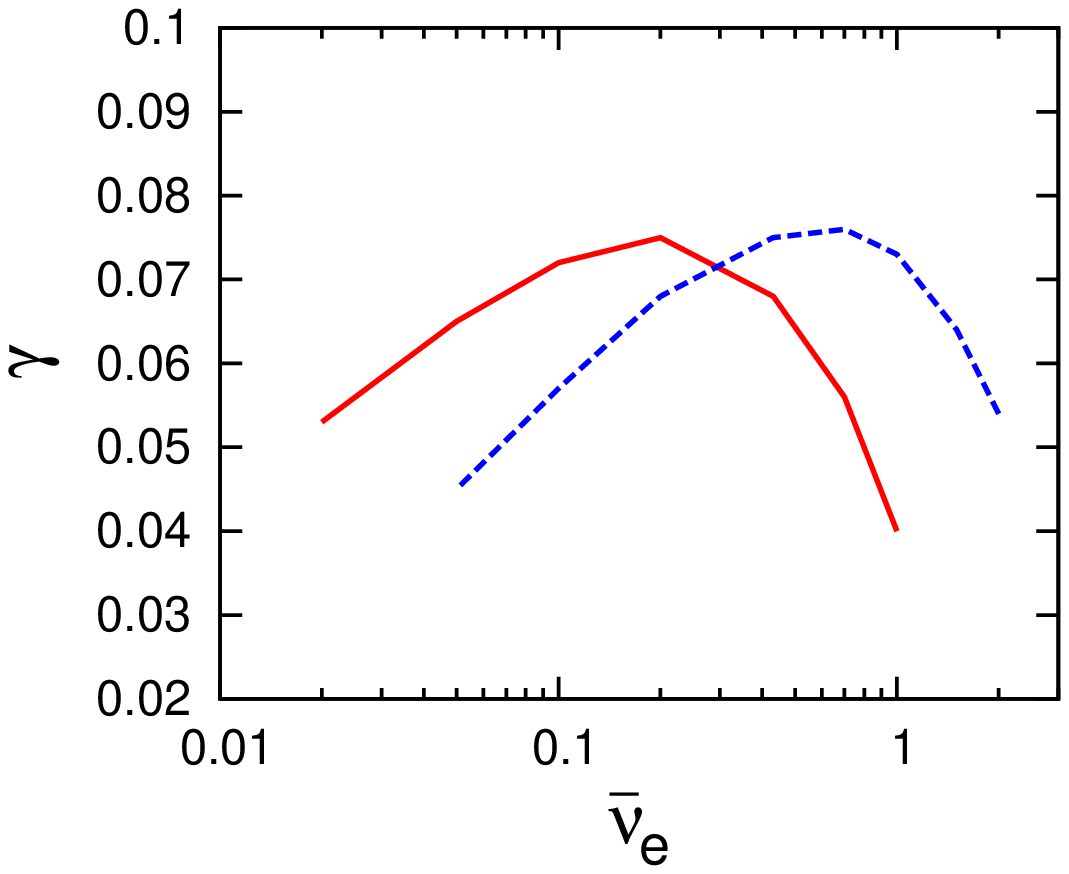}
\includegraphics[angle=270, width=6cm, totalheight=4.7cm,trim=0 0 -20 0,clip]{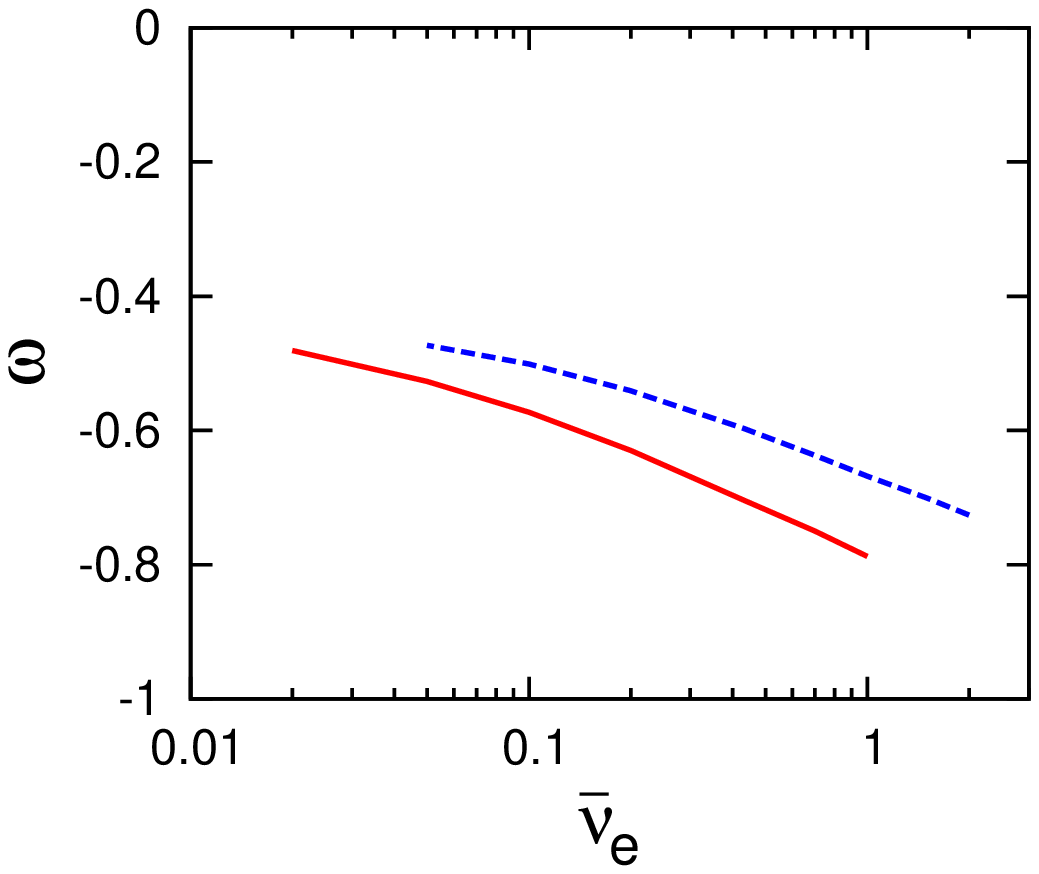}		
\caption{Plots of growth rate and real frequency versus collisionality for calculations with a full ion response (solid lines), and a Boltzmann response for the ions (dashed lines).}
\label{fig:adiabatic2}
\end{figure}

\begin{figure} [tbh]
\centering
\includegraphics[angle=180,width=4.4cm, totalheight=5.25cm,trim=0 0 -50 0, clip]{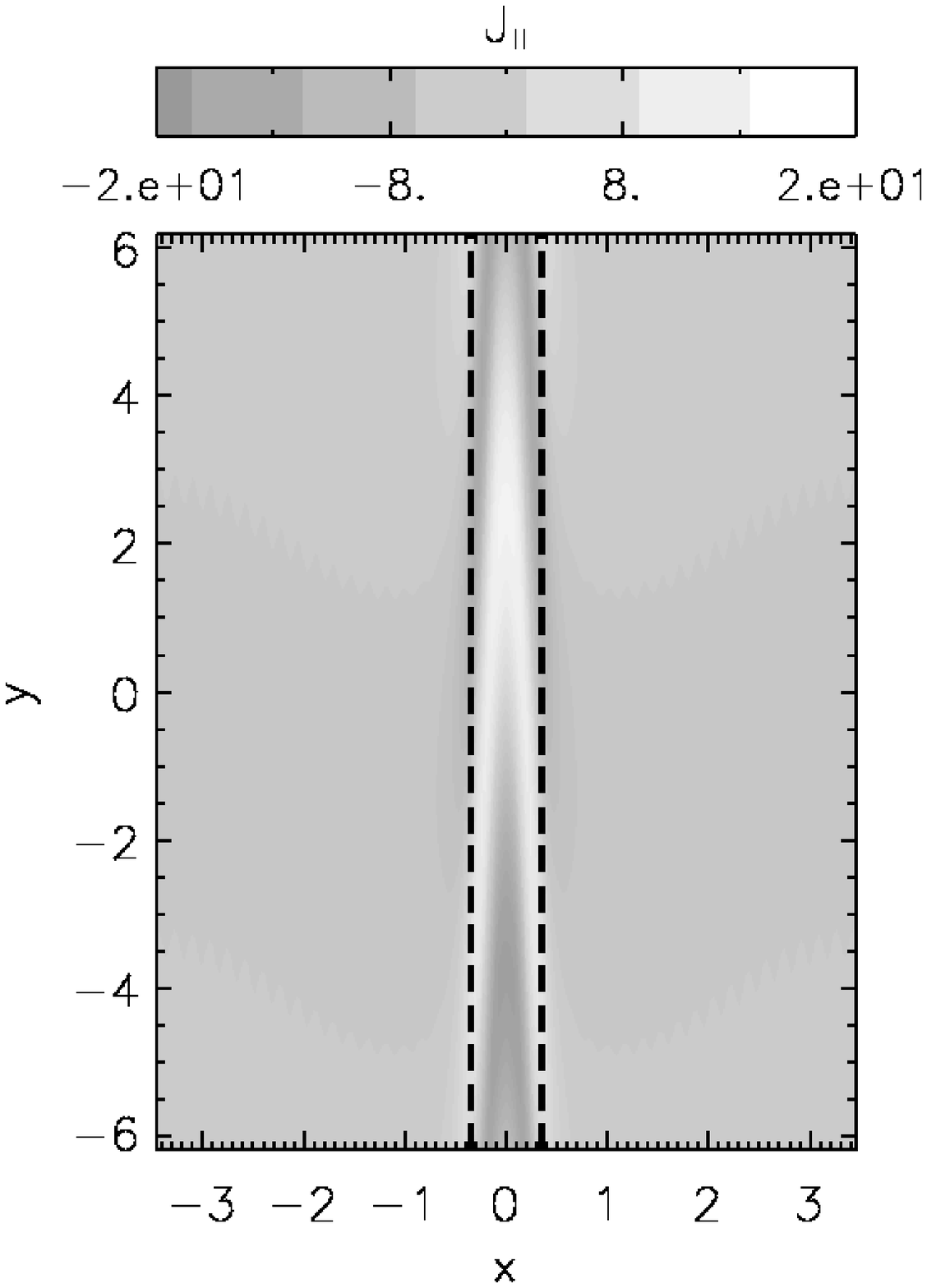}
\includegraphics[angle=270, width=6.7cm, totalheight=5.25cm, trim=-50 -50 -20 0,clip]{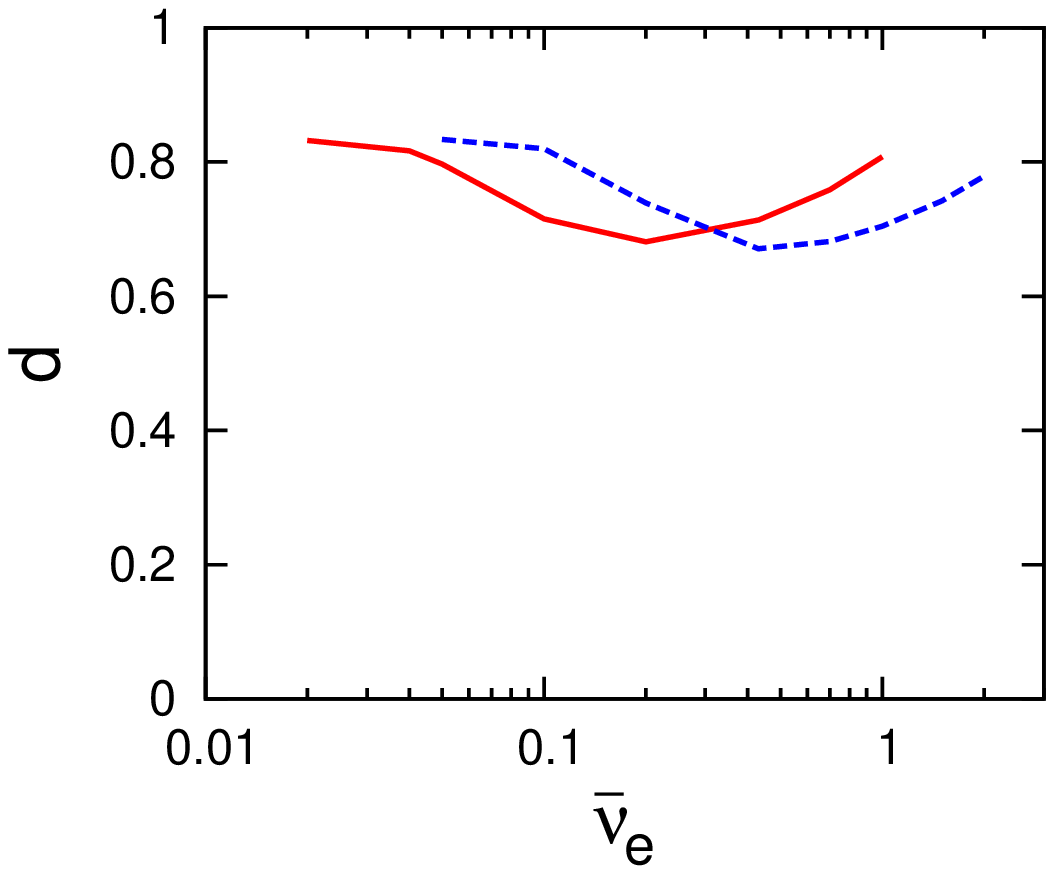}
\caption{(Left) Cross section of the parallel current $\tilde J_\parallel$ in the flux tube at $\theta=0$, and at $k_y \rho_{i}=0.5$. The dashed lines show the width $d$ of the current layer which is defined in equation (\ref{ddefined}). (Right) The current width $d$ is plotted as a function of collisionality $\bar \nu_{e}$, using full ion physics (solid line), and also a Boltzmann response for the ions (dashed line).}
\label{fig:adiabatic3}
\end{figure}

The appropriate form for the ion response in tearing mode calculations has been widely discussed in the literature.  Early calculations by authors such as Drake and Lee \cite{drake}, utilise magnetised or unmagnetised approximations for the ion response, depending on the predicted width of the current layer $d$ (a magnetised response is used if $d\gg\rho_i$ while an unmagnetised response is used for $d \ll \rho_i$). Later calculations by Cowley \emph{et al} \cite{cowley}, implimented the effect of finite ion Larmor radius (FLR) effects more completetly, in the limit $d < \rho_i$.  To discern the effect ion physics has on the MAST micro-tearing instability, a comparison has been made between calculations using the full ion response, and calculations using an unmagnetised ion response,

\begin{equation}
\tilde n_i= -\frac{e \tilde \phi}{T_i}F_0
\end{equation}

\noindent where $\tilde n_i$ is the perturbed ion density.  The results of these calculations are presented in Figure \ref{fig:adiabatic2}, where the growth rate and real frequency of the instability (at $k_y\rho_{i}=0.5$) are plotted against collision frequency.  It is clear the full ion response has a limited influence on the instability, contrasting with the calculations of Cowley \emph{et al} \cite{cowley}, who find FLR effects have a strong stabilising effect on the collisionless and semi-collisional tearing instabilities.  However, this apparent contradiction can be explained by considering the current layer produced by the MAST micro-tearing instability.  Figure \ref{fig:adiabatic3} is a contour plot of the current perturbation produced by the $k_y\rho_{i} = 0.5$ micro-tearing mode at  $\theta=0$.  The current perturbation is concentrated in a small layer around $x=0$, and here we define the width of this layer, $d$, using the following expression,

\begin{equation}
\frac{\int^{d/2}_{-d/2}|\tilde J_\parallel|dxdy}{\int^{\infty}_{-\infty}|\tilde J_\parallel|dxdy}=0.75
\label{ddefined}
\end{equation} 
  
\noindent   Figure \ref{fig:adiabatic3} reveals $d\sim\rho_i$ throughout the unstable collisional range, suggesting the approximation $d<\rho_i$, made by Cowley \emph{et al} \cite{cowley}, is not applicable to the MAST instability.  (In the limit $d<\rho_i$ the current layer itself is largely unmagnetised, but the full ion response is still important outside the current layer.  Indeed it is finite Larmor radius (FLR) effects \emph{outside} of the current layer which Cowley \emph{et al} find has a stabilising effect on the instability.   On the other hand the MAST instability has $d/\rho_i\sim 1$ so ion FLR effects ought to be important both inside and outside of the current layer.)

\subsection{Micro-tearing Instabilities in $s-\alpha$ Geometry}

While numerical equilibria provide a realistic reconstruction of the experimental situation, they are somewhat inflexible when one wishes to modify equilibrium parameters.  Thus in order to study the effect that trapped particles, magnetic drifts and aspect ratio have on  the micro-tearing instability, it is helpful to introduce an analytic equilibrium.  Here we have constructed a shifted-circle, $s-\alpha$ equilibrium \cite{salpha} where the flux surfaces are specified by,

\begin{eqnarray}
R&=&R_0(r)+r \mathrm{cos}\theta \nonumber \\
Z&=&r \mathrm{sin}\theta
\end{eqnarray}

\noindent where $r$ is the minor radius of the flux surface, $R$ and $Z$ are the horizontal and vertical coordinates that specify the flux surface, and $R_0(r)$ is the major radius of the flux surface (the $s-\alpha$ model allows this to be a function of the flux surface label $r$).  In this model the total magnetic drift velocity for electrons is given by,

\begin{equation}
\mbf{v}_R + \mbf{v}_{\nabla B} = -\frac{2v_\parallel^2+v_\perp^2}{2\Omega}\frac{\mbf{R}\times\mbf{B}}{R^2 B} 
\end{equation}

Trapped particles can be included in the $s-\alpha$ model by using the following expression for the variation of the magnetic field strength,

\begin{eqnarray}
B = \frac{B_0}{1 + \epsilon  \mathrm{cos}(\theta)}
\end{eqnarray}

\noindent where $\epsilon=r/R_0$ is the inverse aspect ratio of the flux surface and controls the fraction of trapped particles. 

\begin{figure} [tbh]
\centering
\includegraphics[angle=270, width=6cm, totalheight=4.7cm,trim=0 0 -20 0,clip]{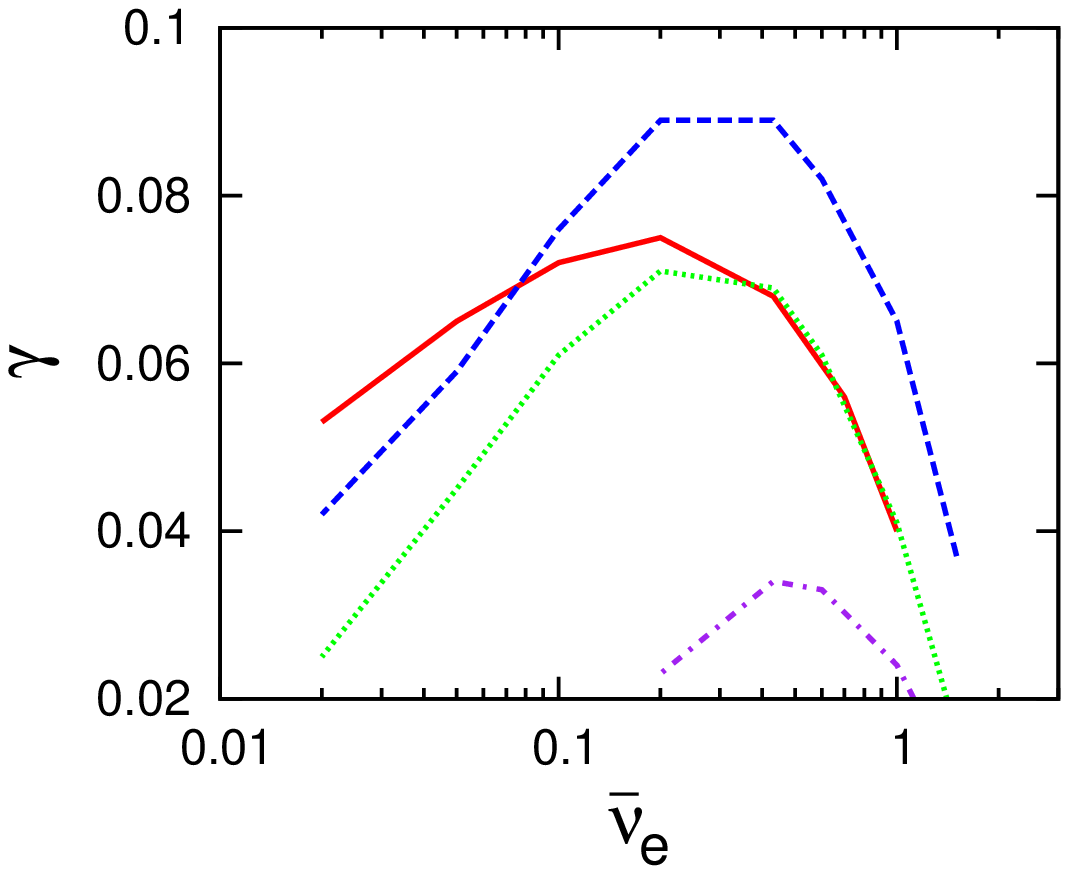}
\includegraphics[angle=270, width=6cm, totalheight=4.7cm,trim=0 0 -20 0,clip]{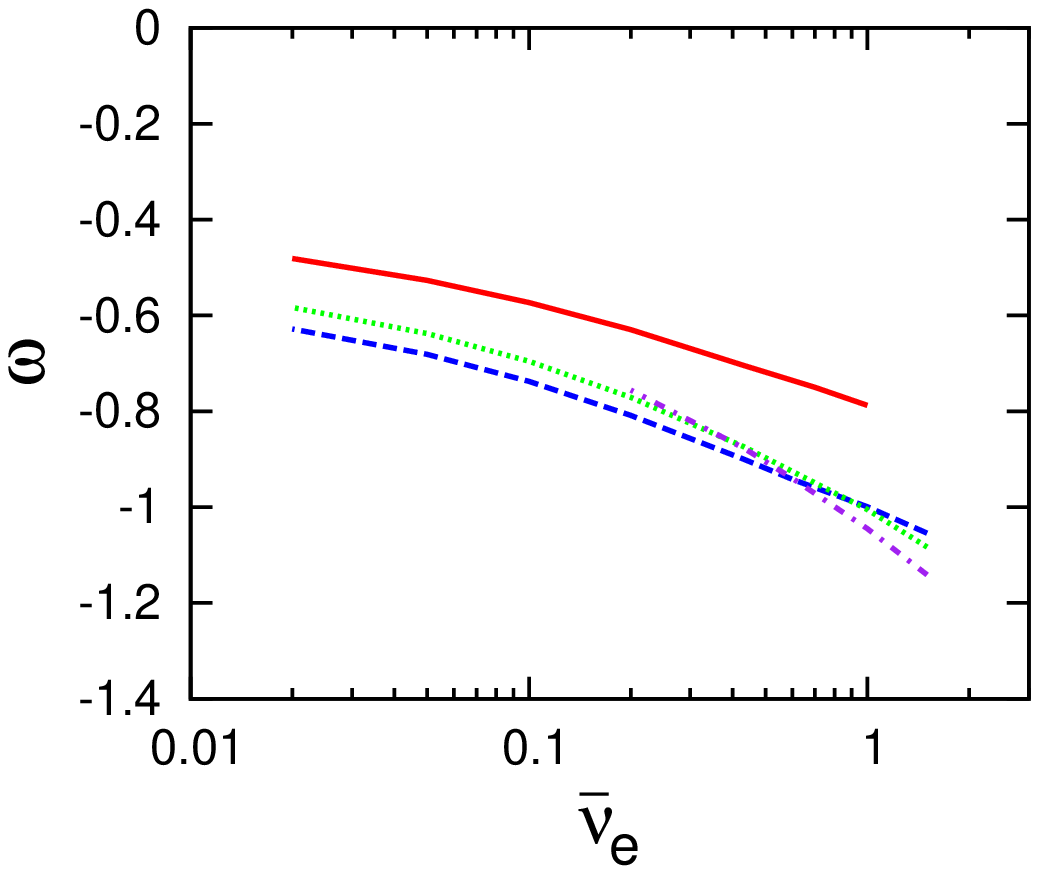}		
\caption{Growth rate and real frequency versus collisionality for four different plasma geometries.  Calculations using the numerical reconstruction are given by solid lines.  Results from the $s-\alpha$ model closest to this equilibrium (using $r/R_0=0.345$ and $R_0=0.879$) are given by dashed lines. The $s-\alpha$ calculations using $r/R_0=0.236$ and $R_0=1.32$ are given by dotted lines. Finally $s-\alpha$ results using $r/R_0=0.118$ and $R_0=2.64$ are given by dash-dot lines.}
\label{fig:salpha}
\end{figure}

In Figure \ref{fig:salpha} the growth rates and real frequencies obtained using the $s-\alpha$ model are compared with those from the numerical MAST equilibrium.  (This is only a rough comparison since the self consistency of the $s-\alpha$ model deteriorates at high $\beta$ \cite{miller}).   The same values of $q$, $\beta$, $\hat s$, $a/L_T$, $a/L_n$, $R_0$ and $r/R_0$ are used in both models.   It is clear the micro-tearing mode is unstable in the $s-\alpha$ equilibrium, suggesting the flux surface shaping of the MAST equilibrium is not particularly important for instability.  

Using the $s-\alpha$ model, the major radius $R_0$ and inverse aspect ratio $\epsilon = r/R_0$ of a flux surface can be altered.  In Figure \ref{fig:salpha}, the value of $R_0$ has been changed keeping $r$ fixed, and allowing $r/R_0$ to vary.  Importantly, all other quantities, such as the temperature and density scale lengths, are kept constant. This is roughly equivalent to changing the aspect ratio $R_0/a$ of a tokamak, whilst staying on the same flux surface $r$.  From Figure \ref{fig:salpha} it is clear that any increase in aspect ratio has a stabilising effect on the micro-tearing instability, although it is also apparent the mode remains significantly unstable if the major radius is modestly increased.  This in turn suggests the instability might appear in moderate aspect ratio machines such as JET (although it will be shown later on that the instability also requires a large $\beta$, which may prevent it from appearing at large aspect ratio).

\subsection{The Effect of Magnetic Drifts} \label{drifts}

In the previous section an $s-\alpha$ model was utilised to vary $R_0$ and $\epsilon=r/R_0$ simultaneously.  Consequently, both the magnetic drift velocity (which is inversely proportional to $R_0$) and the trapped particle fraction (which is given by $\epsilon^{1/2}$) were changed. In this section the $s-\alpha$ model is used  to vary $R_0$ while keeping $\epsilon=r/R_0$ constant, allowing us to consider the effect of magnetic drifts in isolation.  

\begin{figure} [tbh]
	\centering
	\includegraphics[angle=270, width=6cm, totalheight=4.7cm,trim=0 0 -20 0,clip]{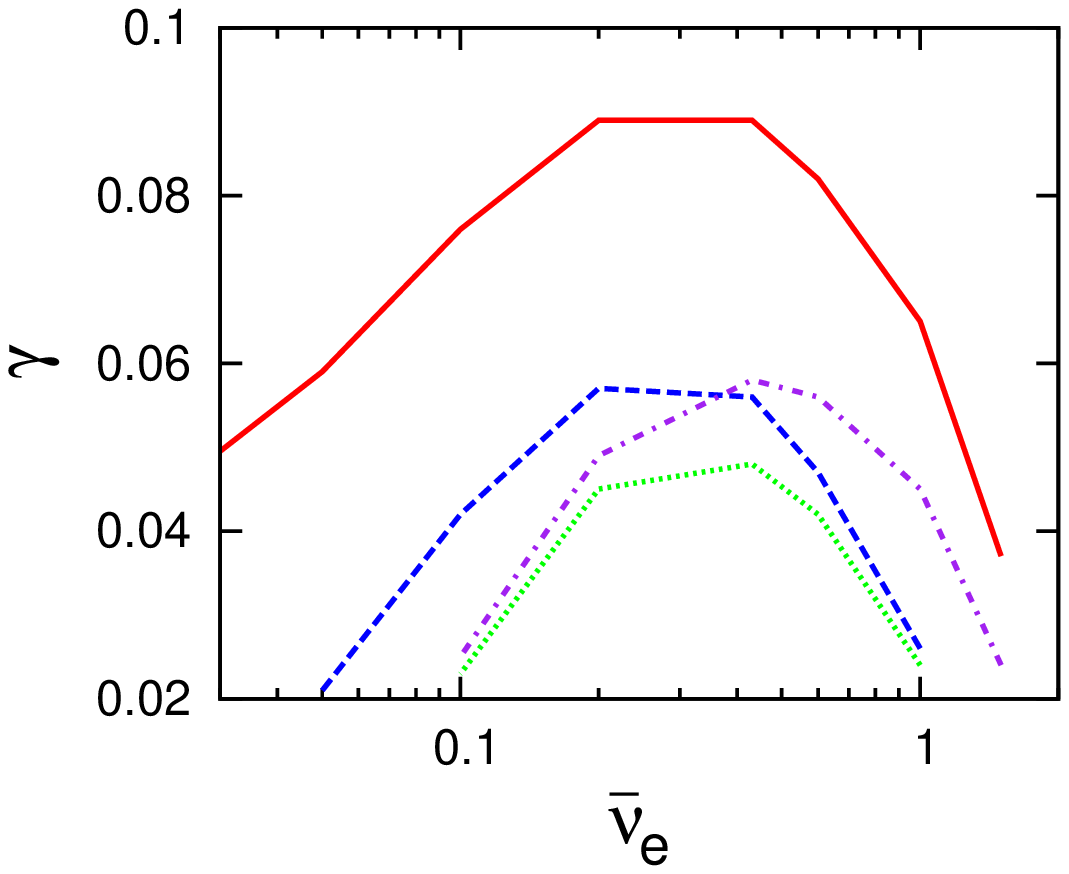}
	\includegraphics[angle=270, width=6cm, totalheight=4.7cm,trim=0 0 -20 	0,clip]{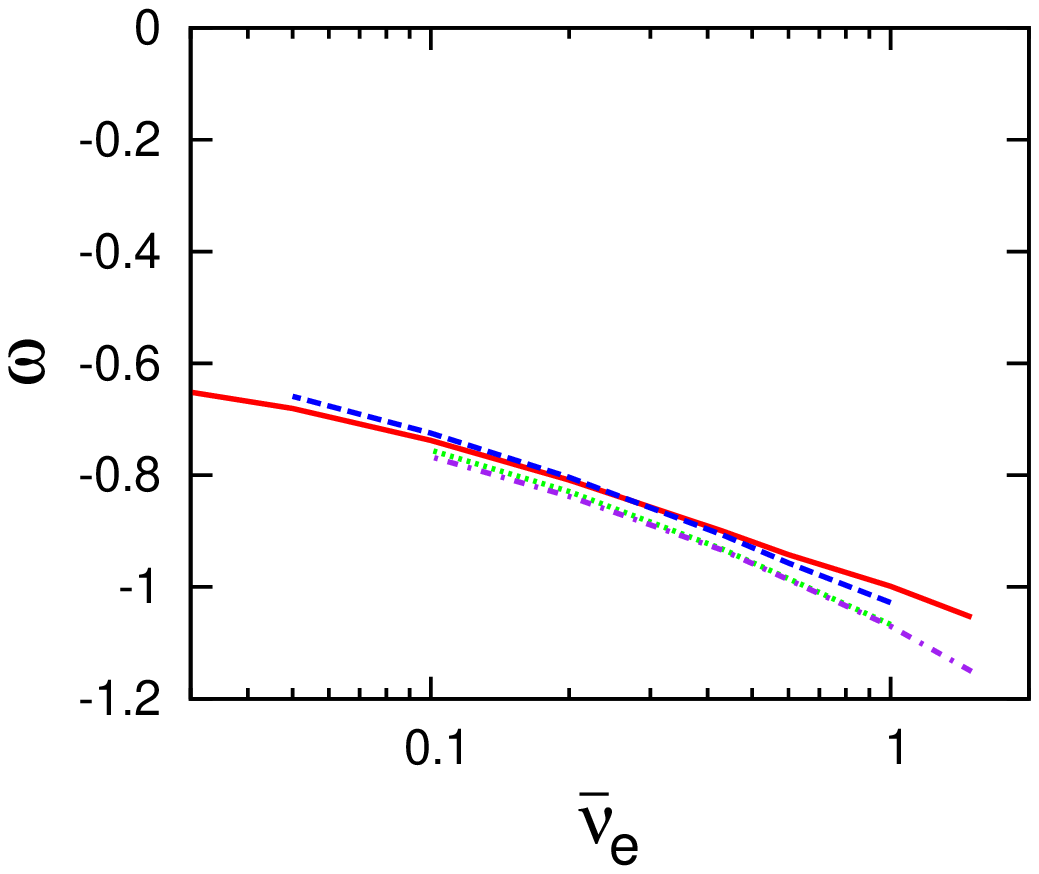}		
	\caption {Plots of growth rate and real frequency versus collisionality using the $s-\alpha$ model with four different major radii at fixed inverse aspect ratio $r/R_0$. Solid lines represent $R_0=0.879$, dashed lines represent $R_0=1.32$, dotted lines represent $R_0=2.64$, dash-dot lines represent $R_0=\infty$ (ie. cylindrical geometry). Note that $k_y \rho_i=0.5$ for all of these calculations.} 
	\label{fig:geo}
\end{figure}

The role that magnetic drifts play in the micro-tearing instability has not been fully treated in the literature. The majority of papers use either slab or cylindrical geometry to help tractability \cite{haz,drake,gladd,cowley}, while a smaller number attempt to understand the role that trapped particles have on the instability \cite{catto,connor}, but ignore the effects that curvature and $\nabla B$ drifts have on untrapped electrons.  Figure \ref{fig:geo} compares the growth rates and real frequencies for calculations with a number of different major radii, and with a range of collisionalities.   These results show magnetic drifts have a significant destabilising effect on the instability, and it is found that the destabilisation comes predominantly via the untrapped electron species. 

It is tempting to suggest the instability is driven by bad curvature at the outboard midplane of the tokamak, which is the destabilisation mechanism for the toroidal ITG/ETG mode.   However, in the case of the micro-tearing mode, the passing electrons must travel many poloidal orbits in an oscillation time of the mode ($\frac{1}{\omega}$), and would be very sensitive to the average magnetic drift rather than just the bad curvature at the outboard midplane.  Also, calculations have been performed where $\alpha=-\frac{2\mu_0 R q^2}{B^2}\frac{dp}{dr}$ has been increased in order to reduce the average curvature (this has been done inconsistently so that the driving gradients are kept constant).  This was found to increase the growth rate of the micro-tearing instability, whereas bad curvature driven modes, such as the toroidal ITG mode, are expected to be stabilised.  Thus, the destabilising influence of magnetic drifts appears to be due to another process (perhaps a resonance between the mode frequency and the magnetic drift frequency as suggested in Section \ref{drift}).

The results from Figure \ref{fig:geo} show there is also an underlying slab drive mechanism for the micro-tearing instability in MAST, since the mode is unstable even when the magnetic drift velocity is zero.  As we mentioned earlier, this additional mechanism appears to be closer to the slab calculations of Gladd \emph{et al} \cite{gladd}, since the instability becomes much more sensitive to the energy dependence of the collision operator in the absence of magnetic drifts.

\subsection{Calculations without trapped particles} \label{trappedeffect}

\begin{figure} [tbh]
\centering
\includegraphics[angle=270, width=6cm, totalheight=4.7cm,trim=0 0 -20 0,clip]{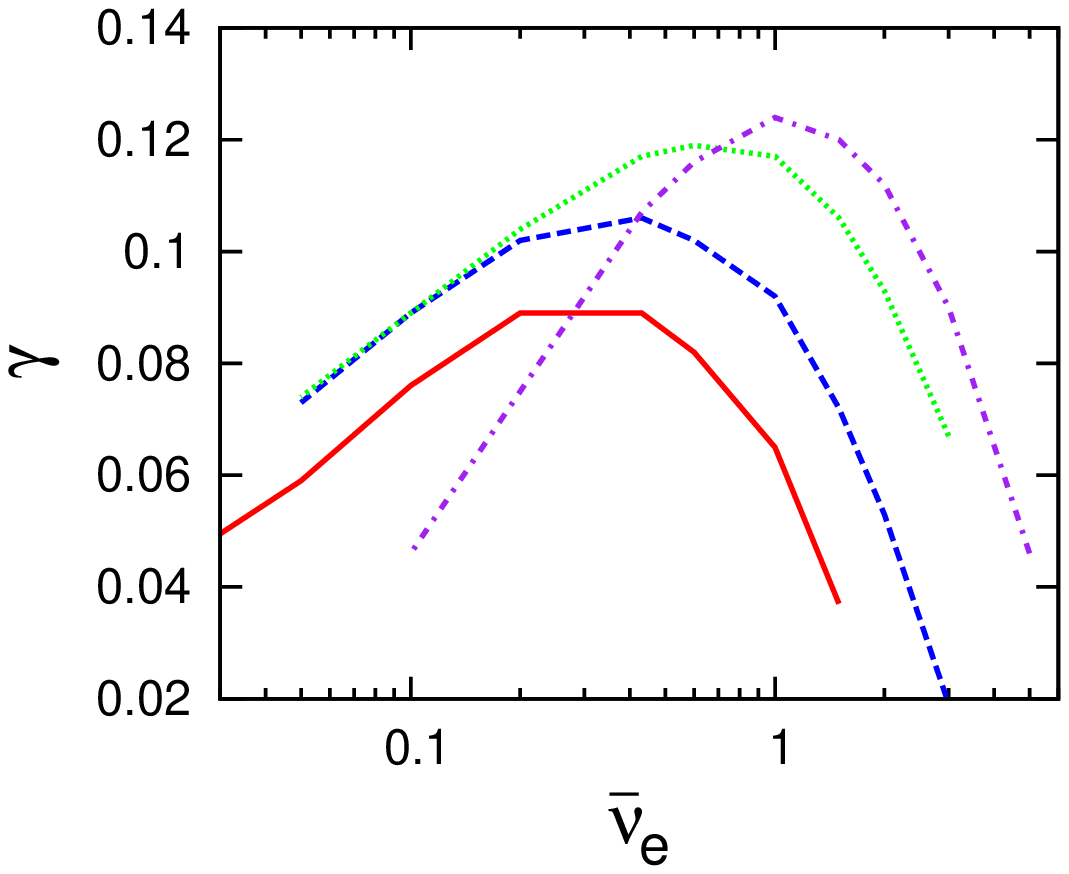}
\includegraphics[angle=270, width=6cm, totalheight=4.7cm,trim=0 0 -20 0,clip]{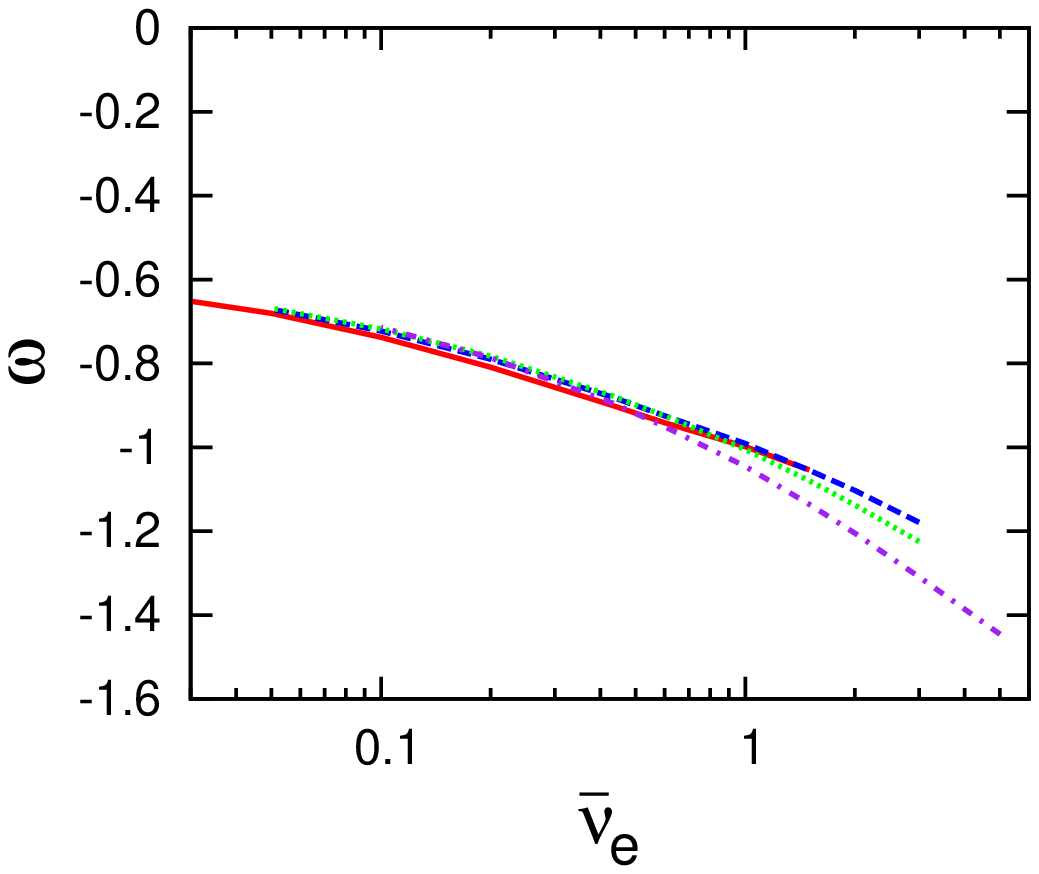}		
\caption{Plots of  growth rate and real frequency versus collisionality using the $s-\alpha$ model with four different inverse aspect ratios $r/R_0$ at fixed major radius $R_0$. Solid lines represent $r/R_0=0.354$, dashed lines represent $r/R_0=0.236$, dotted lines represent $r/R_0=0.118$, and dot-dash lines represent $r/R_0=0$.} \label{fig:trappedresults}
\end{figure}

It is also possible to investigate the role that trapped particles play in the micro-tearing instability using the $s-\alpha$ model.  To do this we simply vary the ratio $r/R_0$ whilst keeping $R_0$ fixed.    Figure \ref{fig:trappedresults} compares the growth rates and real frequencies of calculations with a number of different values of $r/R_0$ and a range of collisionalities.  Importantly these collision frequencies are well below the electron bounce frequency which is $\omega_b=12.6$ (all frequencies are normalised to $\frac{v_{th_i}}{a}$) so the banana orbits of trapped electrons are important.  

Figure \ref{fig:trappedresults} reveals that trapped particles are destabilising at lower collision frequencies, which is in general agreement with the predictions of Catto and Rosenbluth \cite{catto}. It is also apparent that trapped particles are most destabilising at smaller inverse aspect ratio, where the calculations of \cite{catto} are most accurate.  

Importantly the trapped particles have a stabilising effect at higher collision frequencies.  This is probably because the trapped particles cannot respond to a parallel electric field, so in the absence of destabilising boundary layer effects, the trapped particles will significantly reduce the perturbed current in the plasma, and hence damp other drive mechanisms for the micro-tearing instability.  Thus, we can see the stabilising effect becomes stronger as the trapped particle fraction increases, and is particularly strong for the aspect ratio of MAST.

\subsection{The Electrostatic Potential} \label{phieffect}

\begin{figure} [tbh]
\centering
\includegraphics[angle=270, width=6cm, totalheight=4.7cm,trim=0 0 -20 0,clip]{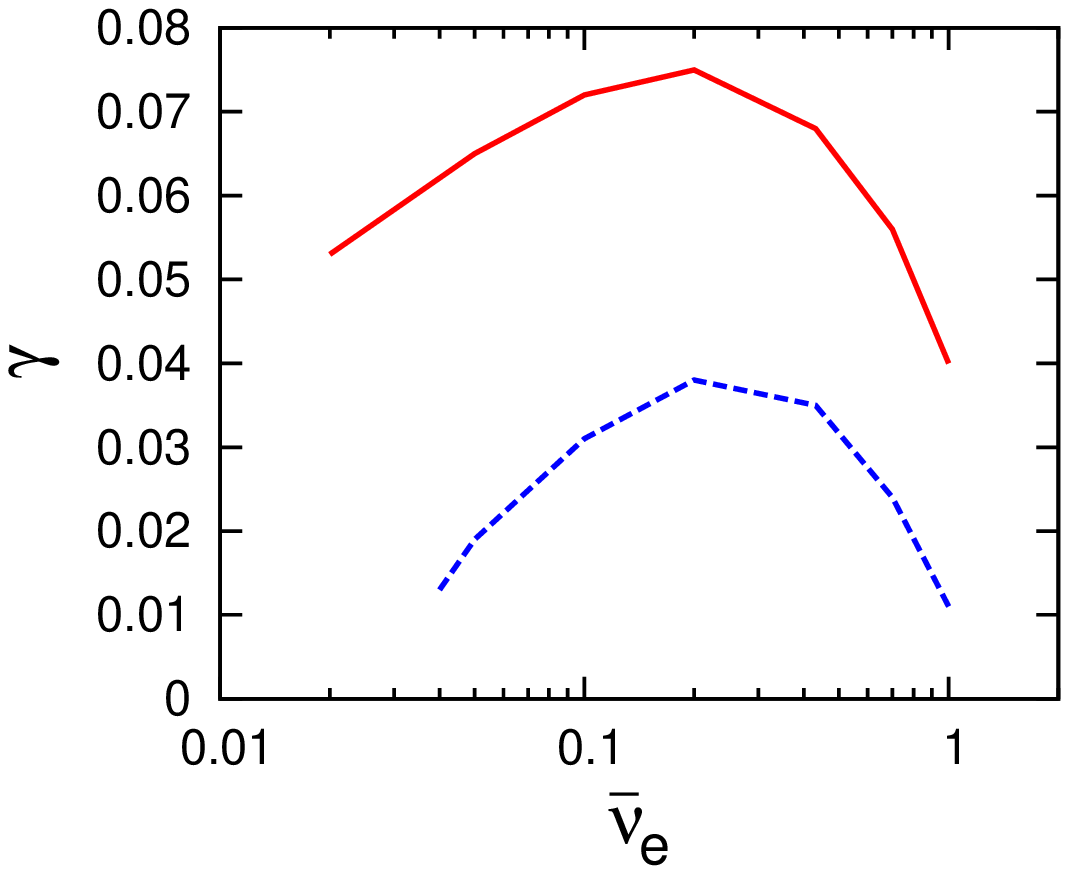}
\includegraphics[angle=270, width=6cm, totalheight=4.7cm,trim=0 0 -20 0,clip]{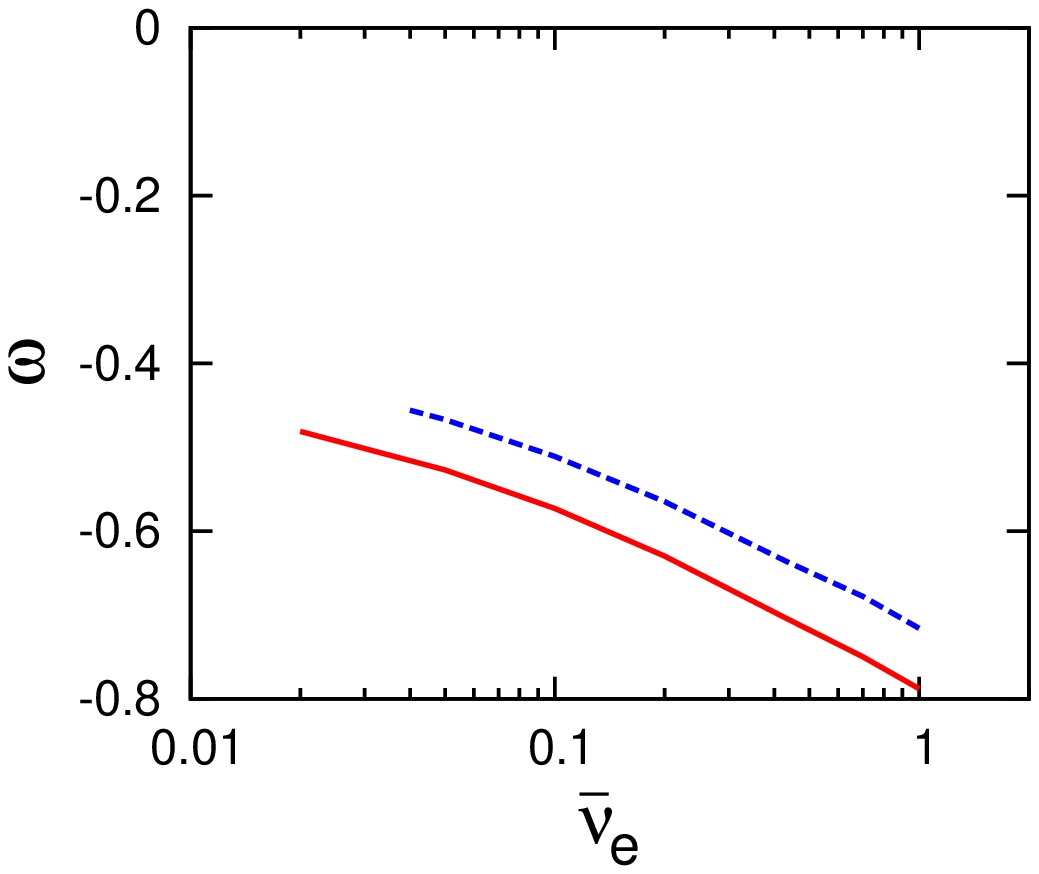}		
\caption{Plots of growth rate and real frequency versus collisionality. Solid lines include $\tilde \phi$, whilst the dashed lines have $\tilde \phi=0$.} \label{fig:phiresults}
\end{figure}

Gladd \emph{et al} \cite{gladd} find the electrostatic potential has a strong destabilising influence on the micro-tearing instability when $\omega \lesssim \bar \nu_{e}$.   The importance of the electrostatic potential has also been tested for the MAST micro-tearing instability by completing calculations with and without the electrostatic potential for a wide range of collisionalities (see Figure \ref{fig:phiresults}).  It is immediately apparent that the electrostatic potential has a strong destabilising influence on the micro-tearing instability. Indeed if both the electrostatic potential and the magnetic drifts are turned off then the mode is stabilised. 

\subsection{The Effect of $\beta$ and $\tilde B_\parallel$} \label{betaeffect}

\begin{figure}[tbh]
\centering
\includegraphics[angle=270, width=6cm, totalheight=4.7cm,trim=0 0 -20 0,clip]{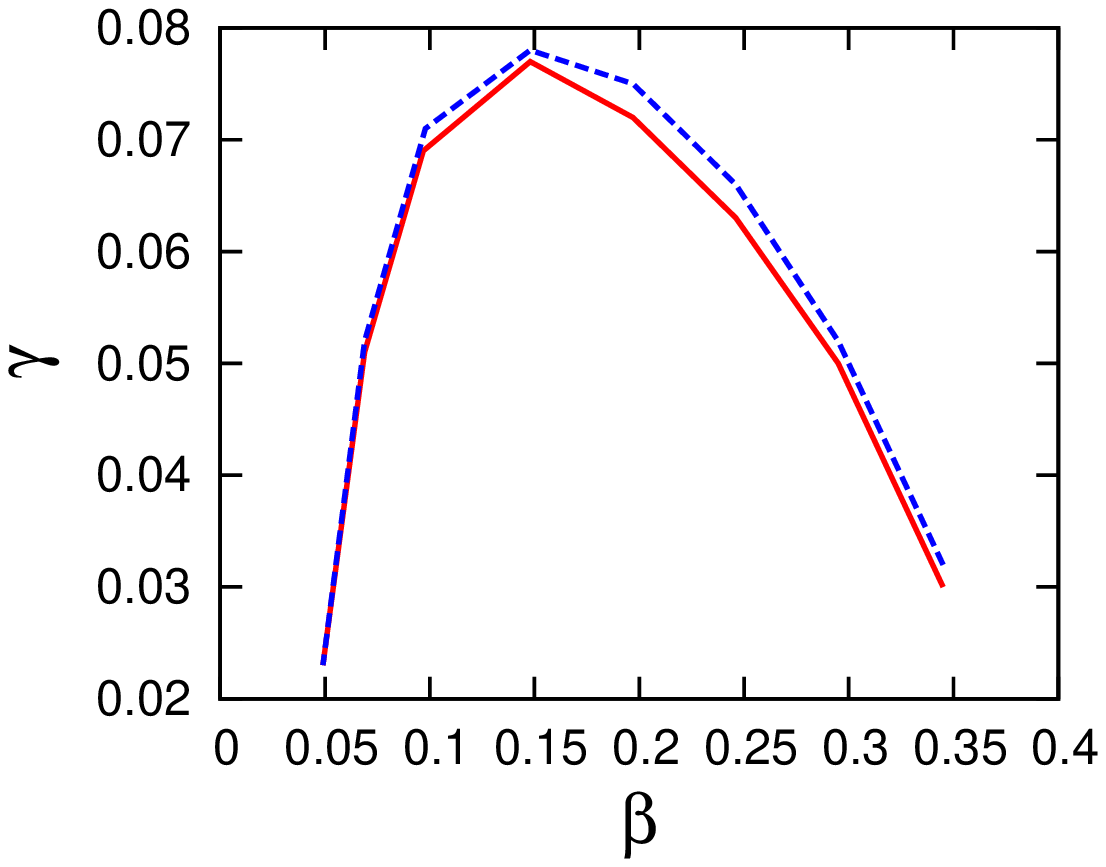}
\includegraphics[angle=270, width=6cm, totalheight=4.7cm,trim=0 0 -10 0,clip]{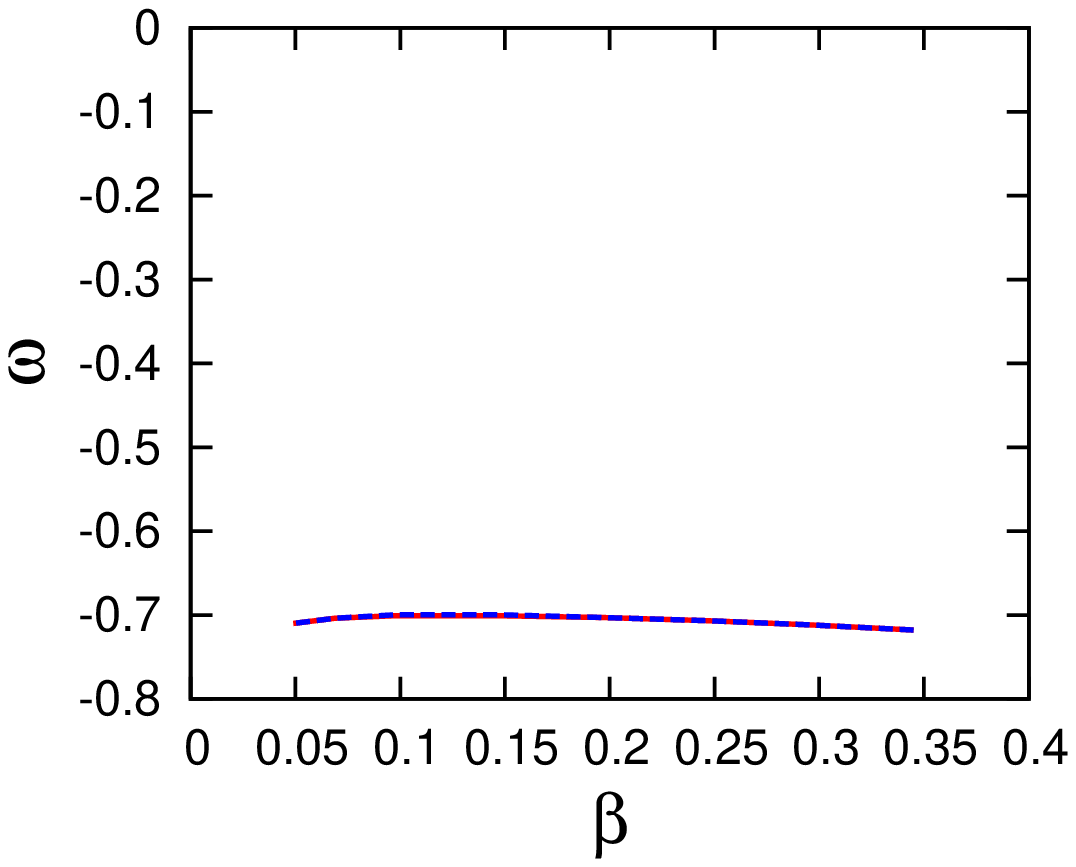}		
\caption{Plots of growth rate and real frequency versus plasma $\beta$. Solid lines include parallel magnetic perturbations, dashed lines do not.}
\label{fig:shat_beta}
\end{figure}

Figure \ref{fig:shat_beta} presents a scan of growth rate versus total $\beta$.  Two models of the plasma are compared, one including parallel magnetic perturbations ($\tilde B_\parallel$) and one without. Importantly, these calculations keep $\beta'$ and $\nabla P$ constant while varying $\beta$ (which is inconsistent), so the strength of magnetic perturbations in Amp\'ere's Law is varied, while the equilibrium is kept the same.    The micro-tearing instability clearly requires a significant $\beta$ value for instability.  Below $\beta \sim 0.1$ (which is the $\beta$ value of the MAST equilibrium) the growth rate decreases rapidly.  This is understandable since the lower values of $\beta$ make it more difficult for the magnetic field to be perturbed.  However, above $\beta \sim 0.15$ the growth rate decreases again, which is not understood at present.    Finally, we note that $\tilde B_\parallel$ is unimportant to the micro-tearing instability, even at $\beta \sim 0.4$.  

\section{Conclusion} \label{conclusion}

The tearing nature of the instability found in MAST has been confirmed by producing Poincar\'e plots of the perturbed magnetic field.  These reveal chains of magnetic islands on the mode rational surfaces, and it is around these rational surfaces that most of the current perturbation lies in a `current layer.' The centre of each island coincides with current minima, while the `X' points coincide with current maxima; and the width of each magnetic island grows as the amplitude of $\tilde A_\parallel$ increases.  These observations are in good agreement with linear tearing mode theory \cite{killeen}.

Previous work describes two different drive mechanisms for the micro-tearing instability.  The first  is due to the thermal force driving parallel currents in the plasma \cite{haz}.  The second is caused by currents flowing in the collisional layers that form around the trapped-passing boundaries, whenever $\bar \nu_{e}/\epsilon<\omega^*_e$ \cite{catto}.  Both of these mechanisms require an energy dependent collision operator, and both are dependent on a finite electron temperature gradient.  Crucially, numerical studies in this paper have shown that neither  drive mechanism is particularly important to the MAST instability, since the mode is largely unaffected when the energy dependence is removed from the collision operator.  The trapped particle drive of Catto and Rosenbluth \cite{catto} appears to have a small destabilising effect when $\bar \nu_{e}/\epsilon<\omega$, but there is clearly another drive mechanism for the micro-tearing instability that has not been described in the literature.  

This new drive mechanism requires an electron temperature gradient from which to draw energy, and has some dependence on the density gradient.  The real frequency of the mode has a roughly linear dependence on both of these gradients, which suggests it has drift wave like properties.  More detailed studies show the micro-tearing mode is strongly destabilised by magnetic drifts and the electrostatic potential.  Indeed, if neither of these effects are included, then the MAST instability is completely stabilised.  

The MAST micro-tearing mode is unstable in the collisional range $0.05 \lesssim \bar \nu_{e}/\omega \lesssim 2$.  The lower end of this range pertains to the calculations of Catto and Rosenbluth \cite{catto} (which have already been mentioned), while the higher end of this range lies somewhere between the collisionless and semi-collisional regimes defined by Drake and Lee  \cite{drake}.  Importantly the collisional range of the MAST instability includes the collisional region $\bar \nu_{e}\sim \omega$, which Gladd \emph{et al} and D'Ippolito \emph{et al} \cite{gladd,dip} have treated numerically. Nonetheless, despite treating the correct collisional range, these papers do not predict the MAST instability since they require an energy dependent collision operator for instability.  However, if we remove magnetic drifts from our calculations then energy dependent collisions do become more important, thus making some contact with these earlier slab calculations \cite{gladd,dip}.

Importantly, it has been found that the exact form of the ion response has little effect on the MAST instability, despite the fact the current layer width $d\sim \rho_i$. Importantly this does not conflict with previous calculations that include the full ion response \cite{cowley}, since these were performed in a different limit ($d<\rho_i$).

Finally, the prevalence of micro-tearing instabilities in simulations of the spherical tokamak is not due to its characteristic shaping, but rather it seems to be a result of the larger $\beta$, and the smaller radius of curvature in the ST (which provides stronger magnetic drifts).

\section{Acknowledgements}

The authors would like to thank A Schekochihin, H R Wilson, M Redi and J F Drake for helpful discussions, and are thankful to N J Conway, A Patel and M J Walsh for contributing data from MAST. This work was funded jointly by the United Kingdom Engineering and Physical Sciences Research Council and by the European Communities under the contract of Association between EURATOM and UKAEA.  The views and opinions expressed herein do not necessarily reflect those of the European Commission.

\appendix

\section*{Appendix: GS2 Coordinates} \label{appendix}

GS2 uses a Clebsch representation for the magnetic field,

\begin{equation}
\mathbf{B}=\nabla \alpha \times \nabla \psi  \label{clebsch}
\end{equation}
 
\noindent where $\psi$ is the poloidal magnetic flux and $\alpha=\phi-q\theta+\nu(\theta,\psi)$ labels different field lines on a flux surface.  Thus, $\psi$ and $\alpha$ are two coordinates perpendicular to the field in the flux tube, while the distance along the field line is labelled using the poloidal angle $\theta$, which extends from $\theta=(-\pi,\pi)$.  In GS2 $\alpha$ and $\psi$ are transformed to $x$ and $y$ using,

\begin{eqnarray}
x=(\psi-\psi_0) \frac{q_0}{a B_a \rho} \label{eqn:xpsi}\\
y=\alpha a \frac{d \psi_{NORM}}{d \rho} \label{eqn:yalpha}
\end{eqnarray}

\noindent where $a$ is the plasma minor radius, $B_a$ is the vacuum magnetic field at the centre of the last closed flux surface, $\psi_{NORM}=\psi/(a^2B_a)$, and $\rho_n$ is a normalised flux surface label increasing from zero at magnetic axis to one at the last closed flux surface.  This results in a coordinate system $(x,y,\theta)$.   

\bibliography{Microtearing_Paper}
\end{document}